\documentclass{emulateapj}

\begin{document}

\author{Nicholas L.\ Chapman\altaffilmark{1},
Shih-Ping Lai\altaffilmark{1,2,3},
Lee G.\ Mundy\altaffilmark{1},
Neal J.\ Evans II\altaffilmark{4},
Timothy Y.\ Brooke\altaffilmark{5},
Lucas A.\ Cieza\altaffilmark{4},
William J.\ Spiesman\altaffilmark{4},
Luisa M.\ Rebull\altaffilmark{6},
Karl R.\ Stapelfeldt\altaffilmark{7},
Alberto Noriega-Crespo\altaffilmark{6},
Lauranne Lanz\altaffilmark{1},
Lori E.\ Allen\altaffilmark{8},
Geoffrey A.\ Blake\altaffilmark{9},
Tyler L.\ Bourke\altaffilmark{8},
Paul M.\ Harvey\altaffilmark{4},
Tracy L.\ Huard\altaffilmark{8},
Jes K.\ J{\o}rgensen\altaffilmark{8},
David W.\ Koerner\altaffilmark{10},
Philip C.\ Myers\altaffilmark{8},
Deborah L.\ Padgett\altaffilmark{6},
Annelia I.\ Sargent\altaffilmark{5},
Peter Teuben\altaffilmark{1},
Ewine F.\ van Dishoeck\altaffilmark{11},
Zahed Wahhaj\altaffilmark{12},
\& Kaisa E.\ Young\altaffilmark{4,13}}

\altaffiltext{1}{Department of Astronomy, University of Maryland, College Park,
MD 20742; chapman@astro.umd.edu}
\altaffiltext{2}{Institute of Astronomy and Department of Physics,
National Tsing Hua University, Hsinchu 30043, Taiwan}
\altaffiltext{3}{Academia Sinica Institute of Astronomy and Astrophysics, P.O.
Box 23-141, Taipei 106, Taiwan}
\altaffiltext{4}{Department of Astronomy, University of Texas at Austin,
1 University Station C1400, Austin, TX 78712}
\altaffiltext{5}{Division of Physics, Mathematics, and Astronomy, MS 105-24,
California Institute of Technology, Pasadena, CA 91125}
\altaffiltext{6}{Spitzer Science Center, MC 220-6, California Institute of
Technology, Pasadena, CA 91125}
\altaffiltext{7}{Jet Propulsion Laboratory, MS 183-900, California Institute
of Technology, Pasadena, CA 91109}
\altaffiltext{8}{Harvard-Smithsonian Center for Astrophysics, 60 Garden Street,
MS42, Cambridge, MA 02138}
\altaffiltext{9}{Division of Geological and Planetary Sciences, MS 150-21, 
California Institute of Technology, Pasadena, CA 91125}
\altaffiltext{10}{Department of Physics and Astronomy, Northern Arizona
University, NAU Box 6010, Flagstaff, AZ 86011-6010}
\altaffiltext{11}{Leiden Observatory, PO Box 9513, NL 2300 RA Leiden, The
Netherlands}
\altaffiltext{12}{Institute for Astronomy, 2680 Woodlawn Drive,
Honolulu, HI~96822}
\altaffiltext{13}{Department of Physical Sciences, Nicholls State University,
Thibodaux, Louisiana 70301}

\title{The Spitzer c2d Survey of Large, Nearby, Interstellar Clouds. IV.
Lupus Observed with MIPS}

\begin{abstract}

We present maps of 7.78 square degrees of the Lupus molecular cloud complex at
24, 70, and $160\:\mu$m. They were made with the Spitzer Space Telescope's
Multiband Imaging Photometer for Spitzer (MIPS) instrument as part of the
Spitzer Legacy Program, ``From Molecular Cores to Planet-Forming Disks'' (c2d).
The maps cover three separate regions in Lupus, denoted I, III, and IV. We
discuss the c2d pipeline and how our data processing differs from it. We
compare source counts in the three regions with two other data sets and
predicted star counts from the Wainscoat model. This comparison shows the
contribution from background galaxies in Lupus I. We also create two color
magnitude diagrams using the 2MASS and MIPS data. From these results, we can
identify background galaxies and distinguish them from probable young stellar
objects. The sources in our catalogs are classified based on their spectral
energy distribution (SED) from 2MASS and Spitzer wavelengths to create a sample
of young stellar object candidates. From 2MASS data, we create extinction
maps for each region and note a strong corresponence between the extinction and
the $160\:\mu$m emission. The masses we derived in each Lupus cloud from our
extinction maps are compared to masses estimated from $^{13}$CO and C$^{18}$O
and found to be similar to our extinction masses in some regions, but
significantly different in others. Finally, based on our color-magnitude
diagrams, we selected 12 of our reddest candidate young stellar objects for
individual discussion. Five of the 12 appear to be newly-discovered YSOs.

\end{abstract}

\keywords{infrared: stars---ISM: clouds---stars: formation}

\section{Introduction}
\setcounter{footnote}{0}

The Spitzer Legacy Science Program, ``From Molecular Cores to Planet-Forming
Disks'' (c2d) \citep{evans03}, mapped five nearby molecular clouds using the
Spitzer Space Telescope (hereafter Spitzer) \citep{werner04}. This paper, fourth
in a series of papers, presents the basic results from the Multiband Imaging
Photometer for Spitzer (MIPS) \citep{rieke04} for the Lupus Molecular cloud
complex.

The Lupus molecular cloud complex covers a broad range of Galactic longitude and
latitude and consists of a number of distinct regions \citep{krautter91}. The
cloud is located near the Scorpius-Centaurus OB association. The large spatial
extent of Lupus on the sky ($\sim20^{\circ}$) means that different regions in
the cloud may be at different distances. Hipparcos measurements to nearby OB
star sub-groups within Scorpius-Centaurus give average distances of $\sim140${}
pc \citep{dezeeuw99}. This distance is in agreement with various reddening
studies \citep[e.g.][]{franco90}. Distance estimates derived from Hipparcos
parallaxes to individual stars within the Lupus clouds can vary by more than a
factor of two, suggesting that depth effects are important in the Lupus
cloud complex \citep{wichmann98,bertout99}. For this paper we are adopting
distances of $150\pm20${} pc for Lupus I and IV and $200\pm20${} pc for Lupus
III. See \citet[and references therein]{comeron06} for a full discussion on the
difficulties in obtaining accurate distance measurements to Lupus.

Compared to other c2d clouds, Lupus is a relatively quiescent star-forming
cloud. \citet{comeron06} has compiled a table of all the known classical T-Tauri
Stars (cTTSs) in different Lupus regions. Within the area we observed with both
the InfraRed Array Camera (IRAC) \citep{fazio04} and MIPS instruments, there are
5 CTTSs in Lupus I, 32 in Lupus III, and 3 in Lupus IV. In addition to the known
CTTSs, \citet{comeron06} lists 19 new suspected very low-mass members of Lupus
III with likely H$\alpha$ emission \citep{lopezmarti05}, and 4 recently
discovered M-type members in Lupus III \citep{comeron03}. Over 100 weak-line
T-Tauri stars (WTTSs) were discovered with ROSAT \citep{krautter97}, extending
over a much larger region than the dark clouds. Based on velocity dispersions
and positions, \citet{wichmann97} speculated that these WTTSs may have formed
earlier and since moved away from the dark clouds.

We summarize our observations in \S\,2 and our data processing pipeline in
\S\,3. A full description of the data processing used on all c2d data is
available in our delivery documentation \citep{evans06}. For this paper, we made
`MIPS high-reliability' cuts to the c2d catalogs, as described in \S\,4. Our
results are summarized in \S\,\ref{sec:results}.

Our goals in this paper are to present the c2d Lupus MIPS data and identify the
likely young stellar members of the Lupus cloud. We plot the differential source
counts at $24\:\mu$m for our cloud regions and compare these to model
predictions and observed off-cloud fields (\S\,\ref{sec:sourcecounts}). We use
the 2MASS and MIPS data to create two color-magnitude diagrams
(\S\,\ref{sec:cmplots}) which are useful in distinguishing background galaxies
from young stellar objects. We create a sample of `Young Stellar Object (YSO)
candidates' based on each source's spectral energy distribution in the 2MASS and
Spitzer wavebands and discuss the significance of our sample in relation to
previously know YSOs (\S\,\ref{sec:yso}). Because YSOs are typically found in
the densest regions, we create extinction maps from 2MASS, estimate the mass in
each cloud region, and compare the extinction maps to the $160\:\mu$m emission
(\S\,\ref{sec:avmaps}). Finally, from our `YSO candidate' sample and the
color-magnitude diagrams, we select 12 likely young stellar objects. These
sources are discussed individually in \S\,\ref{sec:interesting}.

\section{Observations}

\begin{deluxetable}{lccc}
\tablewidth{0pt}
\tablecolumns{4}

\tablecaption{\label{tab:aors} Summary of Observations}

\tablehead{\colhead{Region} & \colhead{AOR Number} & \colhead{Date Observed}
         & \colhead{Program ID}\\
\colhead{} & \colhead{} & \colhead{YYYY-MM-DD} & \colhead{}}
\startdata
Lupus I       & 0005723648 & 2004-08-24 & 175\\
              & 0005722880 & 2004-08-23 & 175\\
              & 0005723904 & 2004-08-24 & 175\\
              & 0005723136 & 2004-08-24 & 175\\
              & 0005724160 & 2004-08-24 & 175\\
              & 0005722624 & 2004-08-23 & 175\\
              & 0005722368 & 2004-08-23 & 175\\
              & 0005723392 & 2004-08-24 & 175\\
Lupus III     & 0005727232 & 2004-08-24 & 175\\
              & 0005728000 & 2004-08-24 & 175\\
              & 0005727488 & 2004-08-24 & 175\\
              & 0005727744 & 2004-08-24 & 175\\
              & 0005728256 & 2004-08-24 & 175\\
              & 0005728512 & 2004-08-24 & 175\\
Lupus IV      & 0005730304 & 2004-08-24 & 175\\
              & 0005730560 & 2004-08-24 & 175\\
Lupus OC1     & 0005734400 & 2004-08-24 & 175\\
              & 0005734656 & 2004-08-24 & 175\\
Lupus OC2     & 0005734912 & 2004-08-24 & 175\\
              & 0005735168 & 2004-08-24 & 175\\
\enddata
\end{deluxetable}

\begin{figure}
\plotone{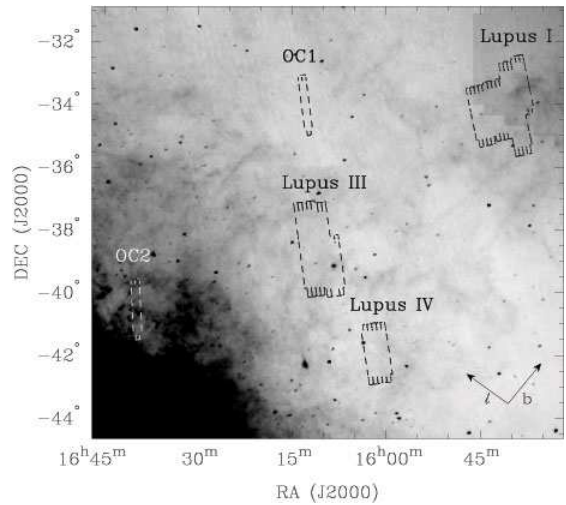}

\caption{\label{fig:iras-map} Map showing the observed regions in the Lupus
cloud. These regions are designated: Lupus I, Lupus III, Lupus IV, and two
off-cloud regions (OC1 and OC2). The outlines show the mapped areas at
$24\:\mu$m; the 70 and $160\:\mu$m coverage maps are very similar. The grayscale
image is the IRAS 25 $\mu$m emission. The saturated region at the lower left is
the Galactic plane.  The directions of increasing Galactic longitude and
latitude are denoted by arrows in the lower right corner.}

\end{figure}

Our observations of Lupus cover three separate on-cloud regions and two
off-cloud regions. We designate the on-cloud regions as I, III, and IV following
the convention from \citet{cambresy99}. The total mapped area on-cloud in two
epochs is approximately 7.78 square degrees. In addition to the on-cloud maps,
we observed two small off-cloud regions for reference, designated off-cloud 1
(OC1) and off-cloud 2 (OC2). In Figure \ref{fig:iras-map}, the observed
$24\:\mu$m areas for each region are overlaid on the $25\:\mu$m IRAS map. The
observed areas at $70$ and $160\:\mu$m are very similar in size, shape, and
position to those at $24\:\mu$m. Table \ref{tab:aors} lists the Astronomical
Observing Request (AOR) numbers, dates observed, and Spitzer program ID for all
our observations.

We used the optical extinction maps of \citet{cambresy99} to choose the areas
mapped. Lupus I covers areas with extinctions $A_V \ge3$, while Lupus III and IV
map extinctions $A_V \ge 2$. The two off-cloud regions had $A_V < 0.5$ in the
optical extinction maps. Lupus II was not observed because there was very little
area in the extinction map with $A_V \ge 2$. Due to time constraints, we did not
observe Lupus V or VI. It is worth noting, however, that there is no obvious
evidence of recent or current star formation in these regions \citep{comeron06}.
Information on the observed regions, including mapped area, distance, and
approximate Galactic coordinates are listed in Table \ref{tab:lup-pars}.

\begin{deluxetable*}{lccccccc}
\tablewidth{0pt}
\tablecolumns{7}

\tablecaption{\label{tab:lup-pars} Basic Information on the Observed Regions}

\tablehead{
\colhead{Region} & \colhead{$l$}    & \colhead{$b$}    & \colhead{dist.} &
\colhead{size}   & \colhead{median $A_V$} & \colhead{Mass ($A_V \ge 3$)}\\
\colhead{}       & \colhead{(deg.)} & \colhead{(deg.)} & \colhead{(pc)} &
\colhead{(deg$^2$)} & \colhead{(mag)} & \colhead{($M_\odot$)}}

\startdata
Lupus I   & 338 & 17 & 150     & 3.82 & 1.8 & 440\\
Lupus III & 340 & 9  & 200     & 2.88 & 2.4 & 690\\
Lupus IV  & 336 & 8  & 150     & 1.08 & 2.2 & 250\\
Lupus OC1 & 344 & 12 & \nodata & 0.31 & 0.7 & \nodata\\
Lupus OC2 & 343 & 4  & \nodata & 0.31 & 2.1 & \nodata\\
\enddata

\tablecomments{ The approximate Galactic latitude and longitude, distance,
mapped area, median $A_V$, and mass are listed for the three on-cloud
regions and the two off-cloud regions. The masses are derived in
\S\,\ref{sec:avmaps}. See \citet{comeron06} for a discussion of the adopted
distances to each region. The map size listed is the size of the $24\:\mu$m
map covered in both epochs. The regions observed at 70 and $160\:\mu$m are
of similar size.}

\end{deluxetable*}

All data were taken in MIPS fast-scan mode with a $240''$ offset between scan
legs. A second epoch of observation was taken later to help identify asteroids.
The time between the two epochs of observation ranged from 3.5 hours in Lupus I
to almost 7 hours in Lupus III. The second epoch is offset, compared to the
first, by $125''$ in the cross-scan direction. Furthermore, to fill in gaps at
$160\:\mu$m, the second epoch is offset by $80''$ along the scan direction. Only
half of the $70\: \mu$m array returns usable data. To minimize this effect, the
AORs were planned such that the second epoch would fill in the holes in the
$70\: \mu$m maps. The net result is that the $24$ and $70\: \mu$m maps have
approximately the same coverage, but the total integration time per pixel at
$70\: \mu$m is 15 seconds, compared to 30 seconds at $24\: \mu$m. The
integration time per pixel is roughly 3 seconds at $160\:\mu$m.

\section{Data Reduction}

The c2d team has made regular deliveries of mosaics and bandmerged catalogs to
the Spitzer Science Center (SSC) for c2d regions. The delivery documentation and
data used in this paper can be found on the SSC website under delivery 4, the
final data delivery of the c2d
team\footnote{http://ssc.spitzer.caltech.edu/legacy/c2dhistory.html}. Detailed
discussion of the data reduction method applied to all c2d products can be found
in the delivery documentation \citep{evans06}. In this paper we will give only a
brief outline of the standard processing procedure. We will discuss the steps of
this procedure in the order they are performed. For this paper, as in delivery
4, we processed all the data with the S13 SSC pipeline.

We made two changes to the standard c2d processing procedure. First, for display
purposes only, we made several cosmetic improvements to the $70\:\mu$m and
$160\:\mu$m images. These enhanced images were created from the older S11 SSC
products. The specific improvements will be discussed in \S\,\ref{sec:image}. 
Source extraction and photometry at these wavelengths is performed with the
standard SSC filtered and unfiltered data products using the c2d pipeline.  Our
second change compared to the standard c2d processing procedure  is that at
$160\:\mu$m we bandmerged the c2d extractions with the c2d catalog.

This paper is one of a series of papers to present the c2d observations of
clouds in a uniform manner. The first paper in this series presented MIPS
results for Chamaeleon II \citep{young05}. That paper used an earlier version of
the SSC pipeline, S9.5. Other papers that present MIPS results from the c2d
clouds use the S13 products as we have here
\citep{rebull07,harvey07a,padgett07}. Despite some differences in data
processing, many of the figures are standardized to facilitate comparisons
between c2d clouds.

\subsection{Image Processing}
\label{sec:image}

\begin{figure*}
\plotone{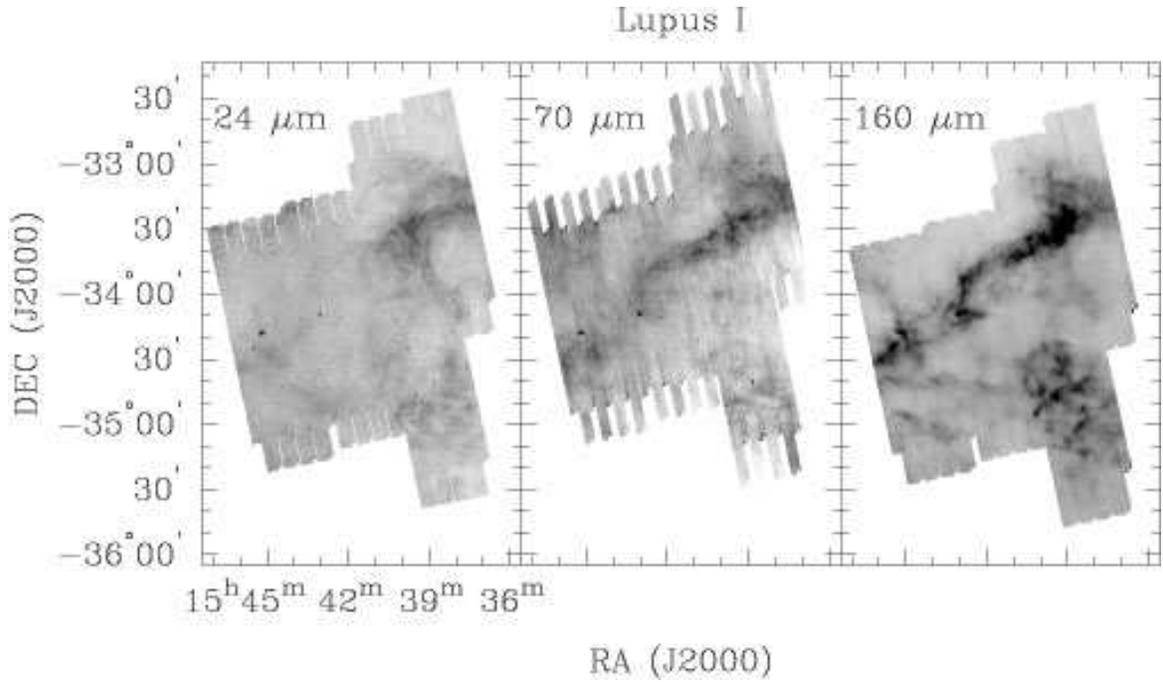}
\caption{\label{fig:lup1-3panel} The 24, 70, and $160\:\mu$m images of Lupus I.
We have applied several cosmetic corrections to the 70 and $160\:\mu$m images
as described in \S\,\ref{sec:image}.}
\end{figure*}

\begin{figure*}
\plotone{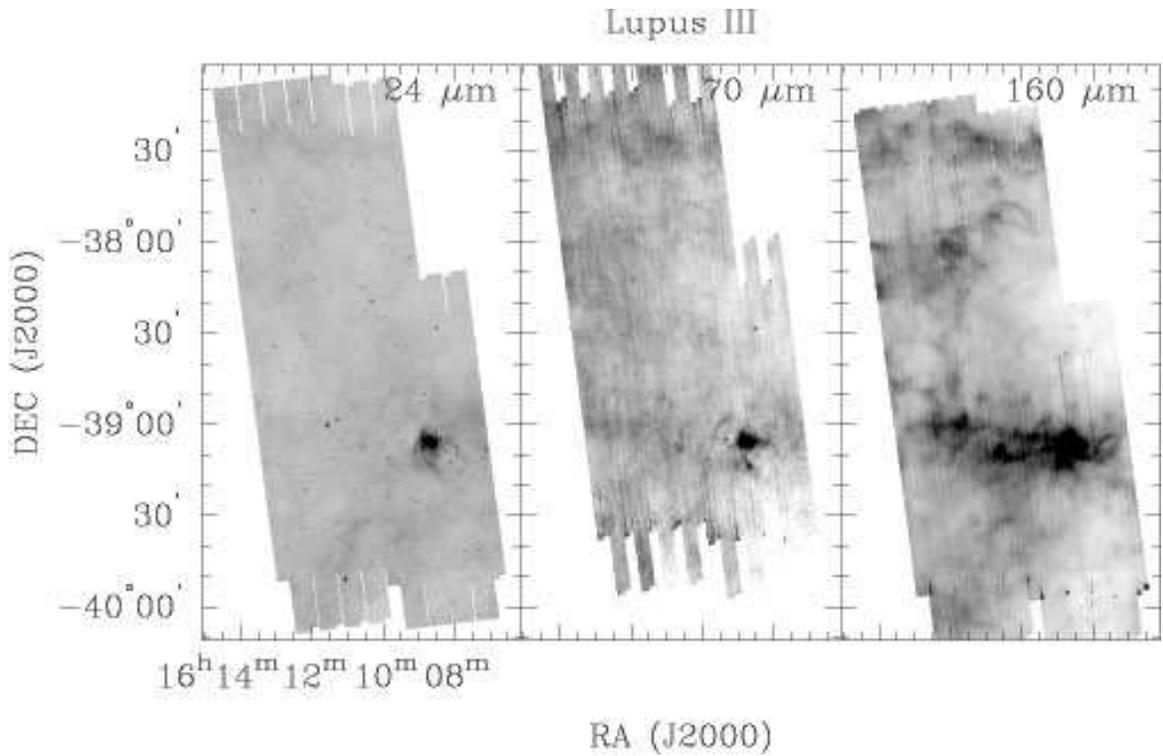}
\caption{\label{fig:lup3-3panel} The 24, 70, and $160\:\mu$m images of Lupus III.
We have applied several cosmetic corrections to the 70 and $160\:\mu$m images
as described in \S\,\ref{sec:image}.}
\end{figure*}

Starting from the Basic Calibrated Datasets (BCDs), the c2d team further
processed the $24\:\mu$m images to remove some instrumental effects. We
corrected a ``jailbar'' response pattern that is caused by bright sources or
cosmic rays by applying an additive correction to the low readouts to bring them
up to the same level as the high readouts. Additionally, the first four frames
in a map were scaled to the median of following frames, and each AOR was
self-flattened to eliminate a small instrumental brightness gradient along the
column direction and other residual image artifacts.

The SSC provided two sets of BCD products for the 70 and 160 $\mu$m data,
unfiltered (normal processing) and filtered (a time median filter is applied).
We have further processed the SSC's unfiltered $70\: \mu$m data to improve the
visual appearance of the mosaic image. The first correction we made was to
subtract out the latent signal from stimulator (stim) flashes. The signal after
the latent had faded was used as a reference to subtract out the stim flash
latents. Further, the SSC unfiltered products contained a striped pattern
roughly along the column direction, which was caused by response variations. We
self-flattened the data with a median obtained from the frames in each scan leg
to reduce this effect.

\begin{figure*}
\plotone{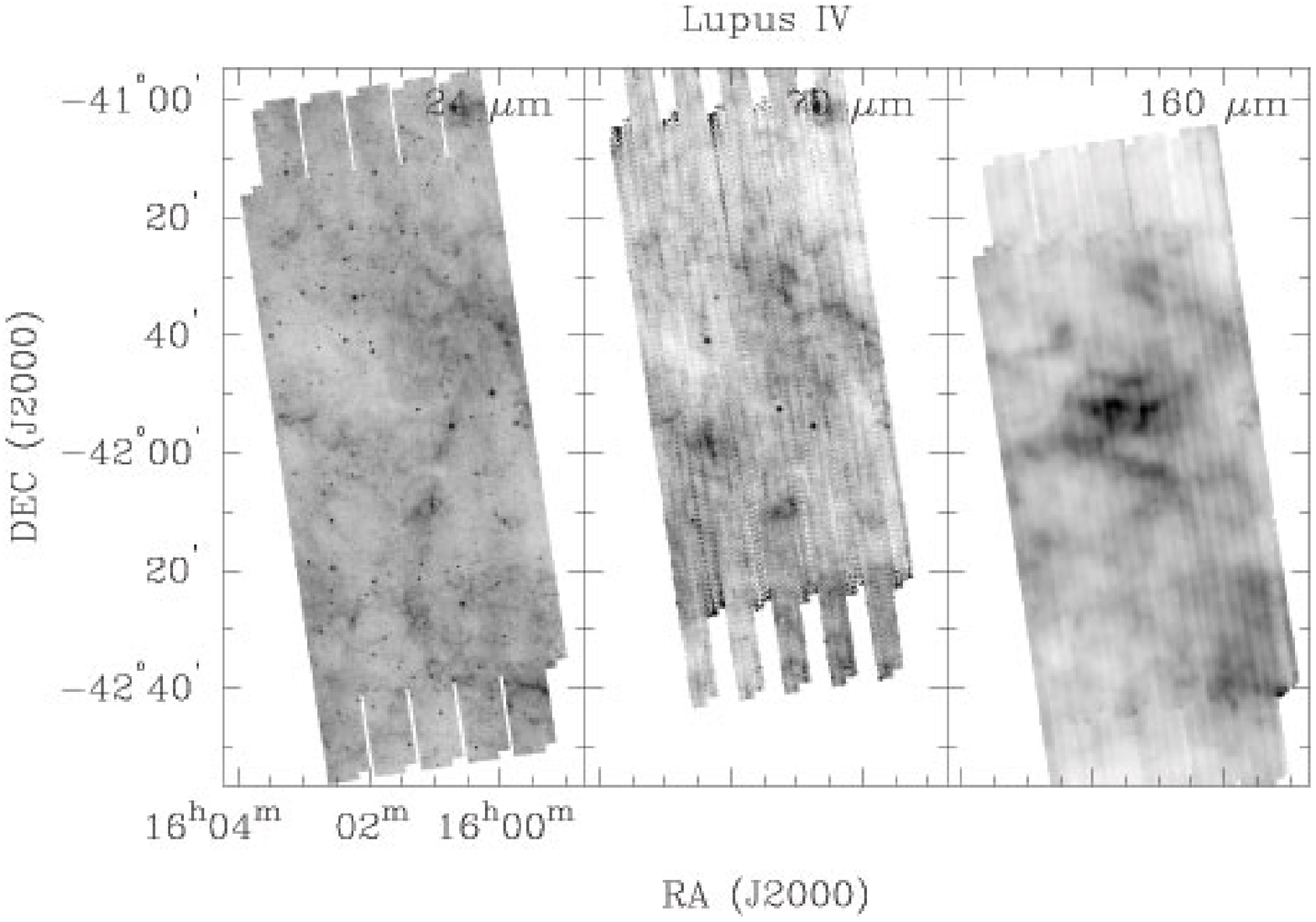}
\caption{\label{fig:lup4-3panel} The 24, 70, and $160\:\mu$m images of Lupus IV.
We have applied several cosmetic corrections to the 70 and $160\:\mu$m images
as described in \S\,\ref{sec:image}.}
\end{figure*}

We have made visually enhanced $160\:\mu$m mosaics by excluding the stim flash
BCDs and the following frames that contained a latent signal. The mosaic was
then created with $8\arcsec$ pixels, half the natural pixel scale (twice the
resolution). Lastly, the image was smoothed with a median filter, where each
pixel in the image was replaced by the median of the values within a $7\times7$
pixel box. Pixels with a value of NaN were treated as missing data during this
filtering.

\begin{figure}
\plotone{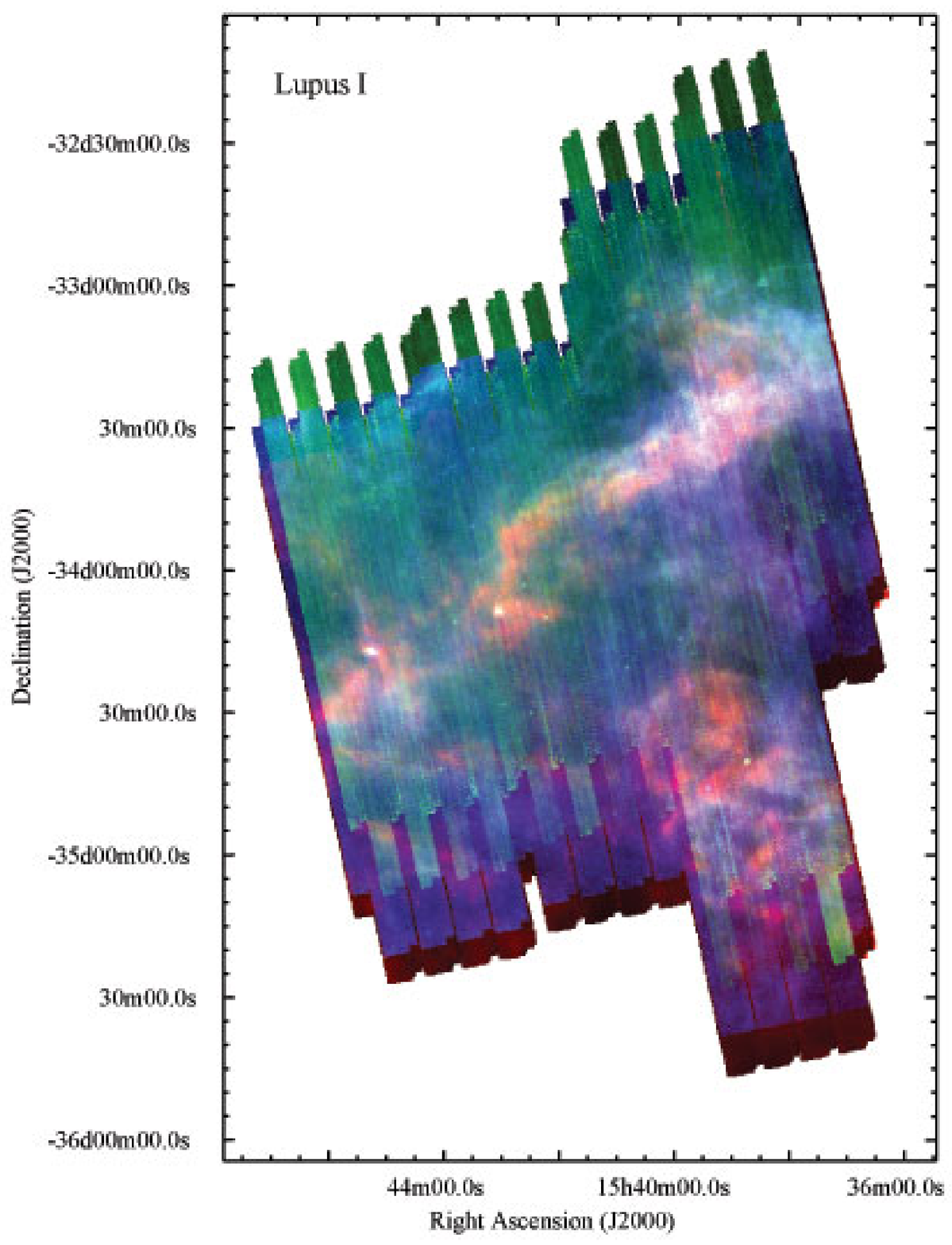}
\caption{\label{fig:lup1-rgb} A pseudo-color image for Lupus I showing the 24,
70, and $160\:\mu$m emission in the blue, green, and red channels, respectively.
We have applied several cosmetic corrections to the 70 and $160\:\mu$m images
as described in \S\,\ref{sec:image}.}
\end{figure}

\begin{figure}
\plotone{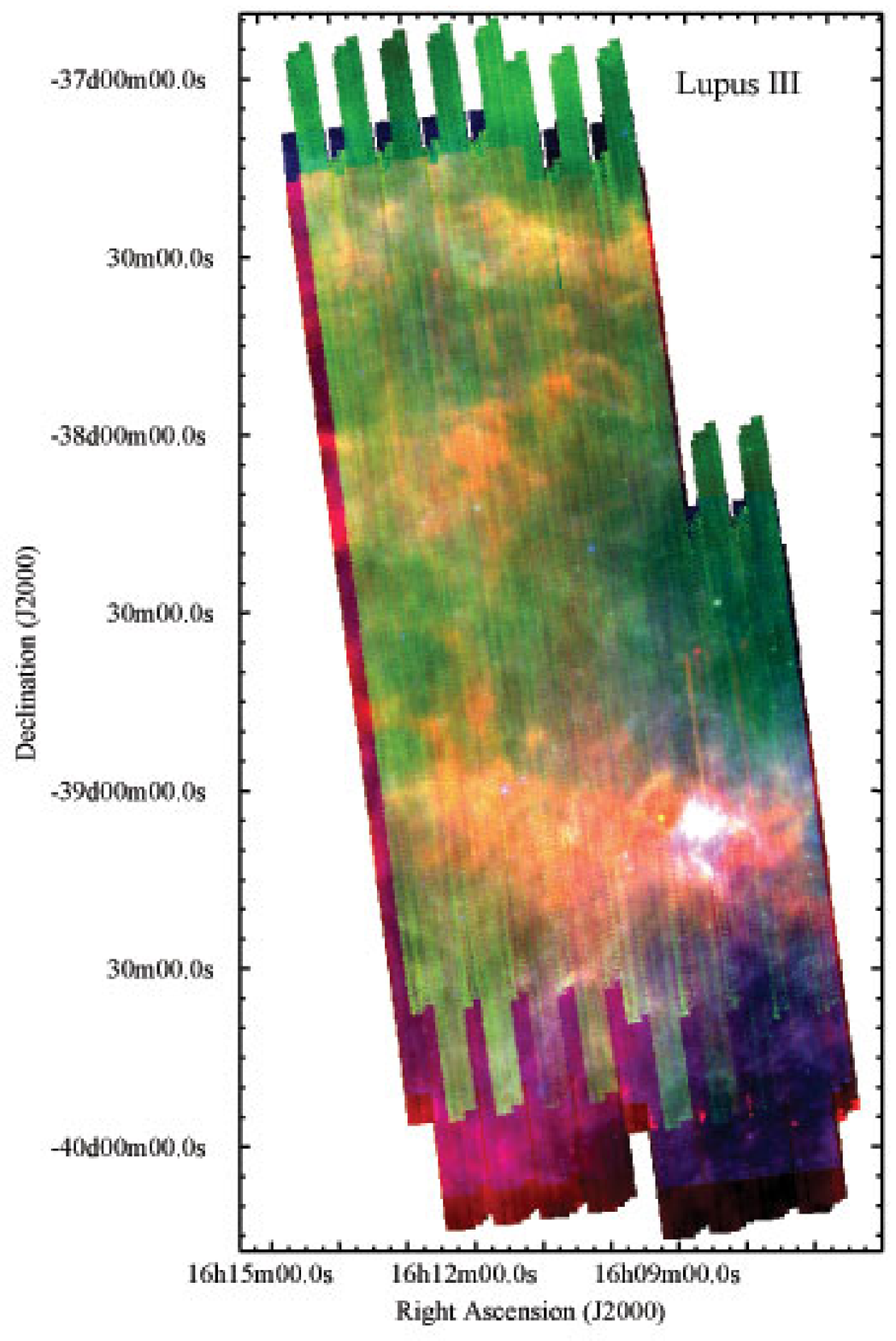}
\caption{\label{fig:lup3-rgb} A pseudo-color image for Lupus III showing the 24,
70, and $160\:\mu$m emission in the blue, green, and red channels, respectively.
We have applied several cosmetic corrections to the 70 and $160\:\mu$m images as
described in \S\,\ref{sec:image}.}
\end{figure}

\begin{figure}
\plotone{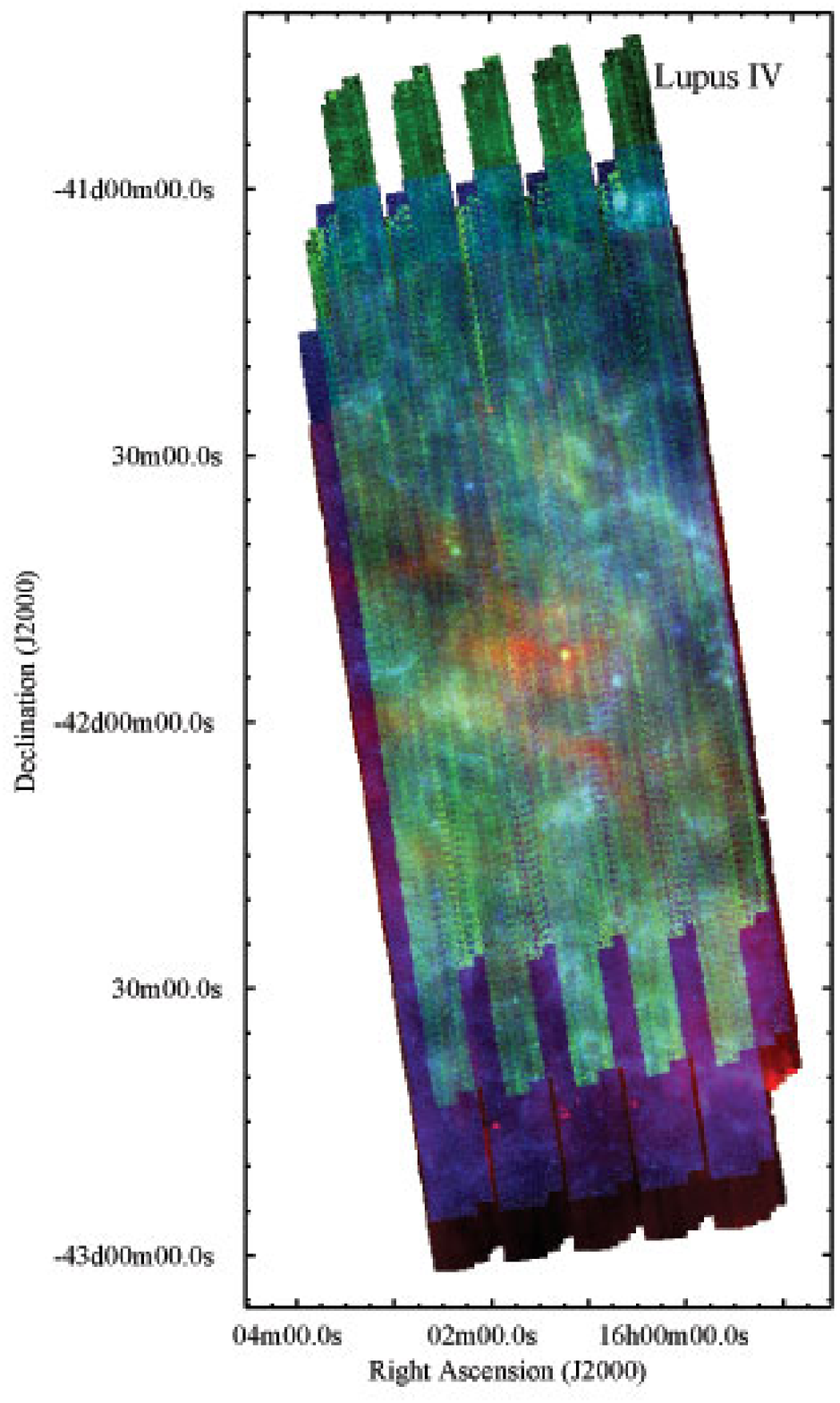}
\caption{\label{fig:lup4-rgb} A pseudo-color image for Lupus IV showing the 24,
70, and $160\:\mu$m emission in the blue, green, and red channels, respectively.
We have applied several cosmetic corrections to the 70 and $160\:\mu$m images as
described in \S\,\ref{sec:image}.}
\end{figure}

Figures \ref{fig:lup1-3panel} - \ref{fig:lup4-3panel} show the 24,70, and 160
$\mu$m images in separate panels for regions I, III, and IV, respectively. We
have also combined these three wavelengths to make a color image where red
represents the $160\:\mu$m, green the $70\:\mu$m, and blue the $24\:\mu$m
emission. The color images for Lupus I, III, and IV are shown in Figures
\ref{fig:lup1-rgb} - \ref{fig:lup4-rgb}.

\subsection{Mosaicking and Source Extraction}
\label{sec:extract}

We created mosaic images at $24\: \mu$m using the SSC's MOPEX program
\citep{makovoz05} with the outlier rejection and position refinement modules.
Three sets of mosaics were created, one for each epoch separately and one
combined epochs mosaic. We used the improved BCDs described in
\S\,\ref{sec:image} for these mosaics. Outlier rejection excluded bad pixels
caused by cosmic rays or image artifacts. The initial identification of point
sources was performed on each mosaic using a modified version of the DOPHOT
program \citep{schechter93}. This program works by fitting point spread
functions, so as sources were found, they were subtracted from the image. This
procedure was iterated until a user defined flux limit was reached. We
calculated the final position and photometry for each source using all BCD
images that contributed to that source.

The $24\:\mu$m source fluxes were calibrated by computing the ratio between the
aperture flux and model point spread function (PSF) flux for all sources with a
reliable aperture flux that could be fitted by the point source profile.  We
assumed the average aperture to PSF flux ratio held for all sources, and a
multiplicative correction was applied.

The $70\:\mu$m point source extractions were performed on mosaic images made
using MOPEX on the SSC's filtered and unfiltered BCDs, \emph{not} the enhanced
mosaics described in \S\,\ref{sec:image}. We used APEX, the SSC's point source
extraction software, on both the filtered and unfiltered mosaic images to obtain
point sources. We used the point response function (PRF) photometry from the
filtered mosaic except for sources brighter than $\sim2$ Jy, when the SSC's
filtering over corrects in the wings of the PRF. In such cases, we substituted
PRF photometry from the unfiltered mosaic.

At $160\:\mu$m we performed source detection and photometry using the MOPEX
package on the SSC's unfiltered BCD mosaics. Again, we did not use the enhanced
mosaics created in \S\,\ref{sec:image}. Fluxes were determined by point-source
fitting. Sources that were close multiples, extended, or had S/N $<5$ are not
listed here. For saturated sources, we assumed the object was a point source and
fit the wings of the PSF to find the flux. Sources which were detected but
deemed unreliable by visual inspection were dropped. Because of this culling,
the 160 $\mu$m source list is believed to be reliable but not complete, and no
completeness limit is given.

The c2d team estimates the absolute flux uncertainty is 15\% at $24\:\mu$m and
$20$\% for $70$ and $160\:\mu$m. We estimate the relative flux uncertainty to be
approximately $5$\% plus the statistical uncertainty for each source.

\subsection{Bandmerging}

The final stage of data processing prior to science analysis of c2d images
involves the creation of bandmerged catalogs. Again, refer to the c2d delivery
document \citep{evans06} for details. Even though this paper focuses on MIPS
results, IRAC bandmerging is also described since these wavelengths are used for
source classification (\S\,\ref{sec:class}) and when discussing individual
sources (\S\,\ref{sec:interesting}).

\subsubsection{IRAC - MIPS 24 Bandmerging}

For each band, the three source extraction lists (epoch1, epoch2, and combined
epochs) were checked for ``self-matches'' within an epoch; two sources extracted
within one epoch, but with positional matches of $\le\: 2\farcs0$ were
considered to be the same source. The fluxes of the detections were summed, and
the position of the source was calculated as the weighted mean. We then merged
the three lists together to cross-identify sources with positional matches of
$2\farcs0$ or less. We visually inspected the epoch-merged source lists for each
band to remove diffraction spikes, column pull-down, latent images, and other
image artifacts that were misidentified as sources.

The epoch-merged source lists for each band were then merged as follows: First,
we merged the four IRAC bands together, one-at-a-time. We started by merging
IRAC1 ($3.6\:\mu$m) and IRAC2 ($4.5\:\mu$m), then merged this product with IRAC3
($5.8\:\mu$m), and finally combined with IRAC4 ($8\:\mu$m). At each step, we
combined detections at two wavelengths into a single source if the difference in
central position was $\le\:2\farcs0$. Then, this bandmerged IRAC catalog was
merged with the $24\:\mu$m MIPS1 band using a larger distance, $4\farcs0$. The
larger distance was used because of the larger PSF at $24\:\mu$m compared to
IRAC. Lastly, the IRAC+MIPS1 catalog was compared with the 2MASS catalog using a
position matching criterion of $2\farcs0$. 

In all steps of bandmerging, IRAC, MIPS1, and 2MASS, we merged together the
closest position match between 2 wavelengths.  Any other detections within the
specified radius were preserved in the final catalog, but not merged.

\subsubsection{MIPS $70\: \mu$m and $160\:\mu$m Bandmerging}

Even though we have two epochs of observation at $70\:\mu$m, they do not have
much overlap between them. Thus, we only performed source extractions on the
combined epochs dataset. First, we performed a `self-merging' on the combined
epochs dataset using a matching radius of $4\farcs0$. Secondly, each $70\: \mu$m
point source detection was matched to all shorter wavelength catalog sources
within a $8\farcs0$ radius. We then visually inspected each merged $70\: \mu$m
detection to determine the `best' source match between shorter wavelengths
($1.25-24\: \mu$m) and $70\: \mu$m. This `best' source match was the one with
the most consistent Spectral Energy Distribution (SED) across all detected
wavelengths. In a few cases where it was not clear which SED was `better', we
chose the closer match.

Shorter wavelength sources that were within $8\farcs0$ of a $70\:\mu$m detection
but \emph{not} matched with the $70\:\mu$m, were assigned a special flag to
denote their status. Thus, in our final catalog, a $70\:\mu$m detection is
listed \emph{once}; we did not try to split up the flux to different shorter
wavelength sources.

There are so few $160\:\mu$m point sources that they were bandmerged
individually by eye for those selected sources described in
\S\,\ref{sec:interesting}.

\subsection{Source Classification}
\label{sec:class}

After bandmerging, we classified sources based on their SEDs at wavelengths from
$1.25-24\: \mu$m. Even though this paper concentrates on the MIPS results for
Lupus, we utilized the IRAC and 2MASS wavelengths for source classification. The
two most important categories for the work discussed here are `star' and `Young
Stellar Object candidate (YSOc)'. Objects classified as `star' have measurements
in 3 or more bands and the SED can be fitted by a reddened stellar photosphere
(including the 2MASS data if available). In addition, if a source had an SED
that could be fitted by a reddened photosphere except in one band, we also
classified it as `star', assuming that it has a prominent spectral line feature
that changed the shape of the SED. A future paper will explore our source
classification method in depth \citep{lai07}.

It is important to include the effect of dust extinction when identifying
stars, because star light is often dimmed by significant dust in the molecular
clouds we observed. The observed flux of a reddened star, $F_{obs}(\lambda)$,
can be described with the following equation:

\begin{equation}
\log(F_{obs}(\lambda)/F_{model}(\lambda)) =
\log(K)-0.4\times C_{ext}(\lambda) \times A_V
\end{equation}

\noindent where $F_{model}(\lambda)$ is the stellar photosphere model, $K$ is
the scaling factor of the model for a particular star, and
\mbox{$C_{ext}(\lambda) \equiv A_{\lambda}/A_V$} is the ratio of extinction at
wavelength $\lambda$ to visual extinction from the dust extinction law. $K$ and
$A_V$ can be derived from the linear fit of this equation by adopting stellar
photosphere and dust extinction models. The stellar photosphere models for
$K_s$--MIPS1 bands are based on the Kurucz-Lejeune models and come from the SSC's
``Star-Pet'' tool\footnote{(http://ssc.spitzer.caltech.edu/tools/starpet)}. For
the 2MASS bands, we translated the observed $J-H$ and $H-K$ colors of stars
\citep{koornneef83} to fluxes relative to $K$ band and ignored the difference
between $K$ and $K_s$ bands. We used a dust extinction model with a ratio of
selective-to-total extinction ($R_V$) of 5.5 \citep[their case
``B'']{weingartner01}. This model is a good match to the \citet{indebetouw05}
extinction law study using the Spitzer IRAC bands.

YSO candidates are traditionally selected from sources with excess flux at near-
to far-infrared wavelengths compared to the stellar photosphere. Unfortunately,
most (if not all) background galaxies fit the same description. In order to
create a sample strongly enriched in YSOs, we have used several statistical
criteria derived from the SWIRE data of ELAIS N1 \citep{surace04} to separate
`YSO candidates' from other objects with infrared excesses. The Spitzer
Wide-area InfraRed Extragalactic (SWIRE) survey imaged six separate fields
selected to be away from the Galactic plane and free from interstellar dust. The
ELAIS N1 region, located towards the Galactic North Pole, presumably contains
almost nothing other than non-YSO stars and background galaxies making it
useful for understanding the distribution of infrared excess non-YSO sources.

We used both IRAC and MIPS data together to identify `YSO candidates'. The
criteria we used were empirically derived from SWIRE and the Serpens molecular
cloud. Detailed discussion can be found in the delivery documentation and
\citet{harvey07b}. In brief, we constructed ``probability'' functions from three
color-magnitude diagrams: [4.5] vs. [4.5] - [8.0], [24] vs. [4.5] - [8.0], and
[24] vs. [8.0] - [24]. Based on where a source lies in each color-magnitude
diagram, it is assigned a probability of being a galaxy. The final probability
is then the product of the three individual probabilities, and if it is less
than a set value, the source is classified as a `YSO candidate'.

\section{Final Catalogs}

All catalogs discussed in the paper are available on the SSC website under the
delivery 4
products\footnote{http://ssc.spitzer.caltech.edu/legacy/c2dhistory.html}.
Separate catalogs of all `YSO candidates' in Lupus are also available from the
same location.

\subsection{Main Lupus Catalogs}

In addition to the visual inspection performed during bandmerging, we made two
further cuts on the data to obtain a high reliability catalog for each cloud
region. First, we required all $24\: \mu$m detections to be at least $3\sigma$
in the combined epochs (quality of detection ``A'', ``B'', or ``C''). Asteroid
contamination is also a problem in Lupus since the cloud is located near the
plane of the ecliptic (latitude -14 to -21 degrees). To eliminate asteroids, we
imposed the requirement that a source be detected at $24\: \mu$m in both of the
single epochs of observation and in the combined epochs data. Because our two
epochs of observation were separated in time by $3-7$ hours, the asteroids would
have moved sufficiently that they would not be detected in both epochs at the
same position. A broader statistical analysis of field asteroids in Spitzer
Legacy datasets will be presented in \citet{stapelfeldt07}. Our 
catalogs contain 1790 sources in Lupus I, 1950 in Lupus III, and 770 in Lupus
IV. In this paper, we only use these `MIPS high-reliability' catalogs.

\subsection{c2d Processed $24\:\mu$m SWIRE Catalogs}
\label{sec:swire}

We have compared our results to the SWIRE data for ELAIS N1 \citep{surace04}.
Because this region maps the Galactic north pole, the sources should be almost
entirely non-YSO stars or background galaxies. Thus, this data set is valuable
in distinguishing YSOs from galaxies. We processed a portion of the SWIRE BCDs
for ELAIS N1 with the c2d data pipeline and created our own c2d-processed SWIRE
catalog. This catalog has only one epoch of observation for MIPS. However, ELAIS
N1 is located at $\sim+70$ ecliptic latitude so asteroid contamination is not a
concern.

The SWIRE data is more sensitive at all wavelengths than our c2d
observations. To make comparisons between SWIRE and c2d, we have created
resampled SWIRE catalogs that simulate the sensitivity of our c2d
observations. This resampling involves: (1) extincting the SWIRE sources
consistent with positioning these sources behind each Lupus region; (2)
randomly selecting faint SWIRE detections to flag as non-detections in the
resampled catalog in order to match sensitivity with c2d; (3) replacing the
photometric errors with ones more characteristic of c2d observations; and
(4) re-classifying the sources in the resampled catalog as described in
\S\,\ref{sec:class}.

After resampling, we have three c2d-processed SWIRE catalogs, one each for Lupus
I, III, and IV. To create `MIPS high-reliability' catalogs similar to above, we
have also required a detection at $24\:\mu$m to be at least $3\sigma$ (quality
of detection ``A'',``B'', or ``C''). Our catalogs contain 1588 sources (Lupus
I), 1646 (Lupus III), and 2061 (Lupus IV). Each one covers 5.3 deg$^2$. We will
use these resampled, `MIPS high-reliability' c2d-processed SWIRE catalogs in
this paper.

\subsection{Lupus Off-Cloud regions}

We observed two off-cloud regions, called OC1 and OC2. For the analysis in this
paper, we will only be using OC1. OC1 is located at a higher Galactic latitude
compared to OC2. The Galactic latitude for OC1 is closer to the latitude of the
three cloud regions. Because of this, the source counts and source
classifications in OC1 provide the best comparison with the on-cloud regions. We
made the same `MIPS high-reliability' cuts to the data as used for the on-cloud
catalogs. Our OC1 catalog contains 157 sources.

\section{Results}
\label{sec:results}

\subsection{$24\:\mu$m Source Counts}
\label{sec:sourcecounts}

\begin{figure}
\plotone{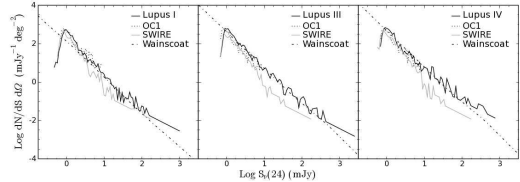}

\caption{\label{fig:sourcecounts} Differential source counts at 24 microns for
the three Lupus regions. Four different curves are displayed: The on-cloud
source counts (black), Lupus off-cloud 1 (dotted), resampled SWIRE (gray), and
25 micron Wainscoat galaxy model counts (dot-dashed) \citep{wainscoat92}. The
Wainscoat model in each panel was computed for the Galactic latitude, longitude,
distance, and average $A_V$ of the specified Lupus region using the values
listed in table \ref{tab:lup-pars}.}

\end{figure}

Figure \ref{fig:sourcecounts} shows the $24\:\mu$m differential source counts
per square degree for Lupus I, III, and IV. For comparison, we also show the
source counts per square degree for Lupus OC1, our c2d-processed SWIRE catalogs,
and the predicted values from the Wainscoat model at 25 microns
\citep{wainscoat92}. The Wainscoat model counts were produced separately for
Lupus I, III, and IV using the Galactic longitude, latitude, distance, and
median $A_V$ present within the each region. These parameters are listed in
Table \ref{tab:lup-pars}. We did not apply a correction for the difference
between the IRAS $25\:\mu$m counts output from the Wainscoat model and our
plotted $24\:\mu$m differential source counts. The source counts for every curve
were binned with a bin width of 0.1 mag., the same bin width produced by the
Wainscoat model.

In all areas, both on- and off-cloud, the counts peak around 1 mJy and
drop off dramatically below this flux due to the completeness limit. Brighter
than $\sim10$ mJy, there are too few $24\:\mu$m sources in OC1 to make a
reliable histogram. Therefore we have not plotted this curve past $10$ mJy.
For fainter fluxes, the source counts for OC1 are consistent with all
three Lupus cloud regions.

The predicted source counts from the Wainscoat model agree well with the actual
on-cloud source counts for Lupus I, III, and IV. The only significant difference
from the model is below $\sim3$ mJy in Lupus I. Here, the on-cloud counts are
higher than those predicted by the Wainscoat model, but are consistent with the
resampled SWIRE counts, suggesting that these sources may be extragalactic. For
Lupus III and IV, the on-cloud counts are slightly higher than the SWIRE counts
at all fluxes. Because Lupus III and IV are consistent with the Wainscoat model,
the higher counts are likely due to Galactic sources. This explanation is
consistent with the relative Galactic latitudes of each cloud. Lupus I, at
$+17$, should have a higher fraction of extragalactic sources compared to Lupus
III and IV which are at lower latitudes, and thus should have a higher fraction
of Galactic sources.

\subsection{MIPS Color-Magnitude Diagrams}
\label{sec:cmplots}

\begin{deluxetable*}{rclcclc}
\tablewidth{0pt}
\tablecolumns{7}

\tablecaption{\label{tab:interesting} Individual YSO Candidates}

\tablehead{\colhead{Index} & \colhead{Region} & \colhead{Source Name} & 
\colhead{RA (J2000)} & \colhead{DEC (J2000)} & \colhead{SSC ID} & \colhead{Class}}
\startdata
 1 & I   & IRAS 15356-3430 & 15:38:48.4 & -34:40:38.2 & SSTc2d J153848.4-344038.2 & I\\
 2 & I   & IRAS 15398-3359 & 15:43:02.3 & -34:09:07.5 & SSTc2d J154302.3-340907.5 & I\\
 3 & I   & 2MASS 15450634  & 15:45:06.3 & -34:17:38.1 & SSTc2d J154506.3-341738.1 & Flat\\
 4 & I   & 2MASS 15450887  & 15:45:08.9 & -34:17:33.7 & SSTc2d J154508.9-341733.7 & II\\
 5 & III & 2MASS 16070854  & 16:07:08.6 & -39:14:07.7 & SSTc2d J160708.6-391407.7 & Flat\\
 6 & III & Sz 102          & 16:08:29.7 & -39:03:11.0 & SSTc2d J160829.7-390311.0 & I\\
 7 & III & IRAS 16051-3820 & 16:08:30.7 & -38:28:26.8 & SSTc2d J160830.7-382826.8 & II\\
 8 & III & IRAS 16059-3857 & 16:09:18.1 & -39:04:53.4 & SSTc2d J160918.1-390453.4 & I\\
 9 & IV  & 2MASS 16011549  & 16:01:15.6 & -41:52:35.3 & SSTc2d J160115.6-415235.3 & Flat\\
10 & IV  & IRAS 15587-4125 & 16:02:13.1 & -41:33:36.3 & SSTc2d J160213.1-413336.3 & I\\
11 & IV  & IRAS 15589-4132 & 16:02:21.6 & -41:40:53.7 & SSTc2d J160221.6-414053.7 & I\\
12 & IV  & 2MASS 16023443  & 16:02:34.5 & -42:11:29.7 & SSTc2d J160234.5-411129.7 & Flat\\
\enddata

\tablecomments{ Names, positions, and classifications of individual YSO
candidates discussed in \S\,\ref{sec:interesting}.  The index numbers
identify the sources in the figures.  Classifications are based on the
criteria of \citet{greene94}.}

\end{deluxetable*}

\begin{deluxetable*}{rrrrrrrrrrrrr}
\tablewidth{0pt}
\tablecolumns{13}

\tablecaption{\label{tab:flux} Fluxes for Selected YSO Candidates}

\tablehead{\colhead{} & \multicolumn{3}{c}{2MASS} & \colhead{} & 
\multicolumn{4}{c}{IRAC} & \colhead{} & \multicolumn{3}{c}{MIPS}\\
\cline{2-4} \cline{6-9} \cline{11-13}
\colhead{Index} & \colhead{$J$} & \colhead{$H$} & \colhead{$K_s$} & \colhead{} &
\colhead{$3.6\:\mu$m} & \colhead{$4.5\:\mu$m} 
& \colhead{$5.8\:\mu$m} & \colhead{$8.0\:\mu$m} & \colhead{} & \colhead{$24\:\mu$m}
& \colhead{$70\:\mu$m} & \colhead{$160\:\mu$m}\\
\colhead{} & \colhead{(mJy)} & \colhead{(mJy)} & \colhead{(mJy)} & \colhead{} &
\colhead{(mJy)} & \colhead{(mJy)} & \colhead{(mJy)} & \colhead{(mJy)} & 
\colhead{} & \colhead{(mJy)} & \colhead{(mJy)} & \colhead{(mJy)}}

\startdata
 1 &    2.15 &    3.84 &    6.60 & &    3.95 &    2.76 &    9.58 &    42.7 & &  114 &    2920 &    4230\\
 2 & \nodata & \nodata & \nodata & &    1.97 &    23.2 &    42.6 &     127 & &  961 &   15300 &   57200\\
 3 &   0.955 &    4.01 &    9.61 & &    14.1 &    15.4 &    16.2 &    19.1 & & 69.9 &     204 & \nodata\\
 4 &    21.1 &    56.7 &    87.4 & &     101 &     104 &    100. &    94.3 & &  137 &     702 & \nodata\\
 5 &    1.30 &    5.73 &    15.8 & &    32.9 &    48.3 &    60.6 &    89.3 & &  196 &     189 & \nodata\\
 6 &    2.39 &    3.54 &    6.19 & &    14.4 &    24.1 &    34.0 &    67.0 & &  347 &     257 & \nodata\\
 7 &    410. &    452. &    342. & & \nodata & \nodata & \nodata & \nodata & &  473 &    2200 &    2210\\
 8 & \nodata & \nodata & \nodata & &   0.255 &    1.00 &   0.985 &   0.549 & & 32.4 &    2610 &    8710\\
 9 &   0.424 &    1.67 &    4.83 & &    8.36 &    9.92 &    8.98 &    7.70 & & 75.9 &    1220 &    5390\\
10 &    2.94 &    2.12 &    1.97 & & \nodata &    2.11 & \nodata &    3.80 & &  533 &     276 & \nodata\\
11 &    1.63 &    3.36 &    4.66 & &    5.10 &    4.87 &    12.6 &    42.1 & &  144 &    1030 &    2340\\
12 &    4.39 &    7.45 &    14.0 & &    31.1 & \nodata &    41.5 & \nodata & &  118 &     269 & \nodata\\
\enddata                                                                       

\tablecomments{ The 2MASS and Spitzer fluxes for the twelve selected
sources in Table \ref{tab:interesting}. The SEDs for these sources are
plotted in Figure \ref{fig:interesting}. The statistical uncertainty in
these fluxes is less than the absolute photometric error. As discussed in
the delivery documentation \citep{evans06} and in \S\,\ref{sec:extract},
the absolute photometric error is estimated to be $15\%$ for $3.6-24\:\mu$m
and $20\%$ at 70 and $160\:\mu$m.}

\end{deluxetable*}

\begin{figure}
\plotone{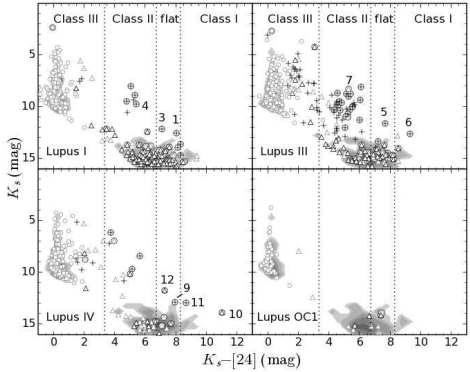}

\caption{\label{fig:lupus-cm} $K_s$ vs. $K_s-[24]$ color-magnitude plot for each
Lupus region and the SWIRE region. The shaded contours are the counts for the
resampled SWIRE region, as described in the text in \mbox{\S\,\ref{sec:swire}}.
The contour levels are 1,2,4,8,16, and 32 counts. The data points show the
specified Lupus region: I, III, IV, or OC1. Different symbols are used based
upon how a source is classified (\S\,\ref{sec:class}). Gray circles are stars,
crosses represent `YSO candidates', gray triangles are sources only detected at
$JHK_s$ and $24\:\mu$m wavelengths, and black triangles are all other
classifications. Furthermore, a black circle surrounds any source also detected
at $70\:\mu$m. The vertical dotted lines show the class boundaries as defined by
\citet{greene94}. The numbers reference sources discussed in
\S\,\ref{sec:interesting} and listed in Tables \ref{tab:interesting} and
\ref{tab:flux}.}

\end{figure}

\begin{figure}
\plotone{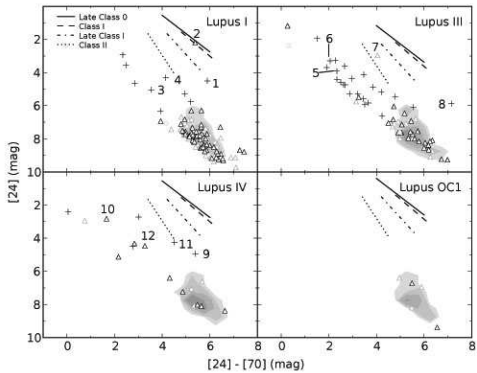}

\caption{\label{fig:lupus-cm-m1m2} $[24]$ vs. $[24] - [70]$ color-magnitude
plots for each Lupus region. See Figure \ref{fig:lupus-cm} for a description of
the symbols, shaded contours, and numbers. Four protostar models are shown from
\citet{whitney03}. Except for Lupus OC1, we shifted the model tracks to
correspond to our assumed distances for each Lupus region.}

\end{figure}

We plotted color-magnitude diagrams for each Lupus region in Figures
\ref{fig:lupus-cm} and \ref{fig:lupus-cm-m1m2}. In each figure, the resampled
SWIRE catalog for the specified Lupus region is shown with shaded contours.
Since we did not create a resampled SWIRE catalog for Lupus OC1, we have shown
the resampled Lupus I SWIRE catalog in the OC1 panels. Overlaid on the contours
are the data points from catalog for each Lupus region. We have plotted the data
using one of four different symbols depending on our source classification: gray
circles are stars, crosses are `YSO candidates', gray triangles denote sources
outside of the observed IRAC area, and black triangles represent all other
categories. Furthermore, in Figure \ref{fig:lupus-cm} we have drawn a black
circle around any source also detected at $70\:\mu$m. A few sources have been
identified by number. These will be discussed in \S\,\ref{sec:interesting}. Note
that source numbers 2 and 8 are missing from Figure \ref{fig:lupus-cm} because
those sources do not have a detection at $K_s$ but do appear in Figure
\ref{fig:lupus-cm-m1m2}.

Figure \ref{fig:lupus-cm} shows the $K_s$ vs. $K_s-[24]$ plot for each cloud
region and OC1. We have drawn dotted lines to separate different YSO classes, as
defined by \citet{greene94}. The four YSO classes are determined by the value of
$\alpha = \frac{\mathrm{d}\log\lambda F_\lambda}{\mathrm{d}\log \lambda}$, the
slope of the linear least-squares fit to all data points from $K_s-24\:\mu$m.
The four classes are as follows: Class I: $\alpha \ge 0.3$; Flat spectrum: $0.3
> \alpha \ge -0.3$; Class II: $-0.3 > \alpha \ge -1.6$; and Class III: $\alpha <
-1.6$. We have included sources with $\alpha = 0.3$ into class I since these
sources were left undefined by \citet{greene94}. We expect the SWIRE data should
contain nothing but non-YSO stars and background galaxies (see
\S\,\ref{sec:swire}). Based on the shaded contours, we can group the sources
into roughly one of three regions in this figure: stars are located at $K_s-[24]
\approx 0$, galaxies are fainter than $K_s \sim 13.5$ mag with $K_s - [24]$
between 4-10, and YSOs have a $K_s - [24]$ color excess and are brighter than
$K_s \sim 13.5$ mag.

We have also plotted $[24]$ vs. $[24] -[70]$ for each region (Figure
\ref{fig:lupus-cm-m1m2}). Only the brightest stars should be detected at
$70\:\mu$m and in fact we do see just one star, in Lupus III, with $[24]-[70]
\approx 0$. Based on the SWIRE contours, we expect that sources with $[24]
\gtrsim 6$ mag are likely galaxies, while those brighter than this limit are
YSOs. We plotted four YSO model curves from \citet{whitney03} showing different
stages of evolution. Each curve represents a range of inclinations from pole-on
(upper left) to edge-on (lower right). These models assume a distance of $140${}
pc, to correspond with nearby star-forming clouds such as Taurus. We shifted the
$24\:\mu$m magnitudes in each sub-panel to correspond to our assumed distance
for that cloud region. We did not apply a shift to the off-cloud region. With
only a few notable exceptions, our `YSO candidates' are bluer than what is
predicted by the models.

\subsection{Young Stellar Objects}
\label{sec:yso}

\citet{comeron06} has compiled the most up-to-date listing of all known and
suspected YSOs in Lupus. Within our observed IRAC+MIPS areas, there are just 5
such objects in Lupus I, 55 in Lupus III, and 3 in Lupus IV. The number of `YSO
candidates' identified from our source classification mirrors this distribution;
there are 16 `YSO candidates' in Lupus I, 75 in Lupus III, and 12 in Lupus IV.
Compared to other c2d clouds such as Perseus, Ophiuchus, and Serpens, Lupus is
relatively quiescent; only of one the regions, Lupus III, has significant star
formation activity. The number of `YSO Candidates' per square degree is 12, 58,
and 33, for Lupus I, III, and IV, respectively. Compared to these values, the
number of `YSO candidates' per square degree is $\sim300$ in Serpens
\citep{harvey06}, $\sim100$ in Perseus \citep{jorgensen06}, and $\sim200$ in the
L1688 region of Ophiuchus \citep{allen07}.

\begin{deluxetable}{lccc}
\tablewidth{0pt}
\tablecolumns{9}

\tablecaption{\label{tab:yso} YSO Candidate Statistics}

\tablehead{\colhead{} & \colhead{Lupus I} & \colhead{Lupus III} &
\colhead{Lupus IV}}

\startdata
YSO candidates      & 16 &  75 & 12\\
YSOs per deg$^2$    & 12 &  58 & 33\\
Known YSOs ($4\arcsec$)\tablenotemark{a} & 3 of 5 & 27 of 55 & 2 of 3\\
\hline
Class I             &  5 &   2 &  1\\
Flat Spectrum       &  2 &   5 &  1\\
Class II            &  6 &  45 &  5\\
Class III           &  3 &  23 &  5\\
\enddata

\tablenotetext{a}{Number of matches with known and suspected YSOs (compiled from
\citealt{comeron06}) within 4 arcseconds.}

\tablecomments{ Statistics on the number of `YSO candidates' identified in
the three Lupus regions based on IRAC and MIPS. See \S\,\ref{sec:class} for
a discussion of how these sources are defined.}

\end{deluxetable}

We have compared our `YSO candidate' sample with the list of known and suspected
YSOs to see how many of the sources we identify. Our results are summarized in
Table \ref{tab:yso}. Overall, we find a match within $4\arcsec$ for 32 out of
the 63 known and suspected YSOs. If we only consider previously known classical
T-Tauri stars, we identify 26 out of 40, or about two-thirds. Of the 31
``missed'' YSOs, 25 are not in our high-reliability catalogs while the remaining
six are not identified as `YSO candidates'. In Table \ref{tab:yso} we have also
broken down our `YSO candidates' by YSO class. For all three cloud regions, the
majority of our objects are in the Class II and Class III categories. This
result agrees with previous studies that have found a number of T-Tauri stars
within the cloud, but no evidence of a large number of Class I and Flat spectrum
stars \citep[e.g.][]{krautter91}.

\begin{figure}
\plotone{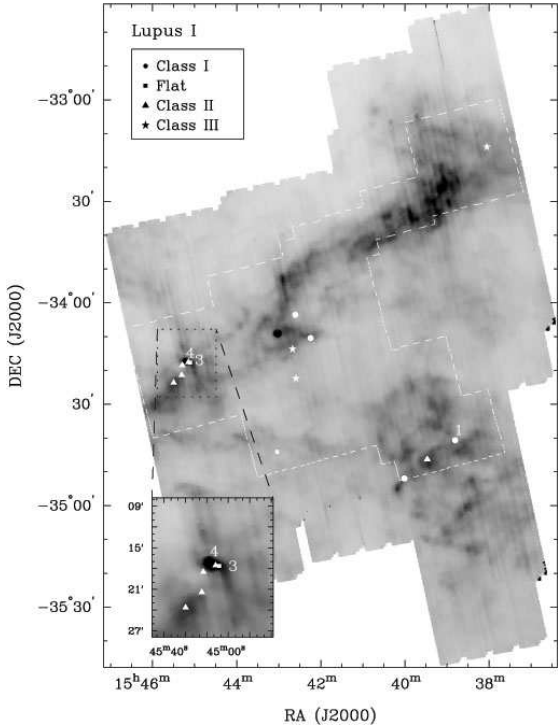}

\caption{\label{fig:lup1-yso}Lupus I `YSO candidates' overlaid on the
$160\:\mu$m image. The four symbols correspond to different YSO classes as noted
in the legend and defined in \S\,\ref{sec:cmplots}. A few sources have been
numbered; these are discussed in \S\,\ref{sec:interesting}. Since our YSO
classification procedures are based on IRAC and MIPS data, we have drawn a
white outline to illustrate the area with both IRAC and MIPS observations.}

\end{figure}

\begin{figure}
\plotone{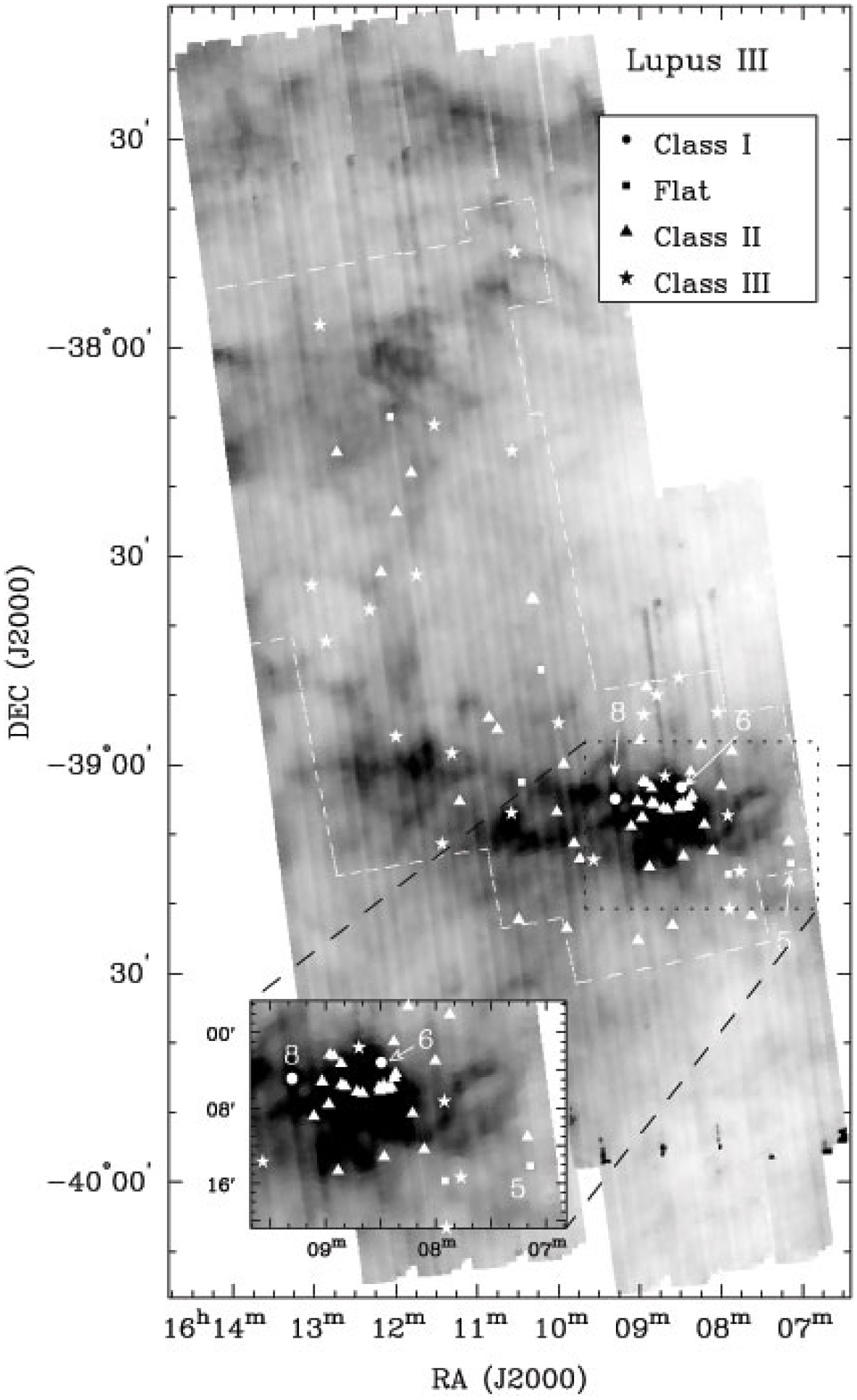}

\caption{\label{fig:lup3-yso}Lupus III `YSO candidates' overlaid on the 
$160\:\mu$m image. See Figure \ref{fig:lup1-yso} for a description of the 
symbols and their meanings.}

\end{figure}

\begin{figure}
\plotone{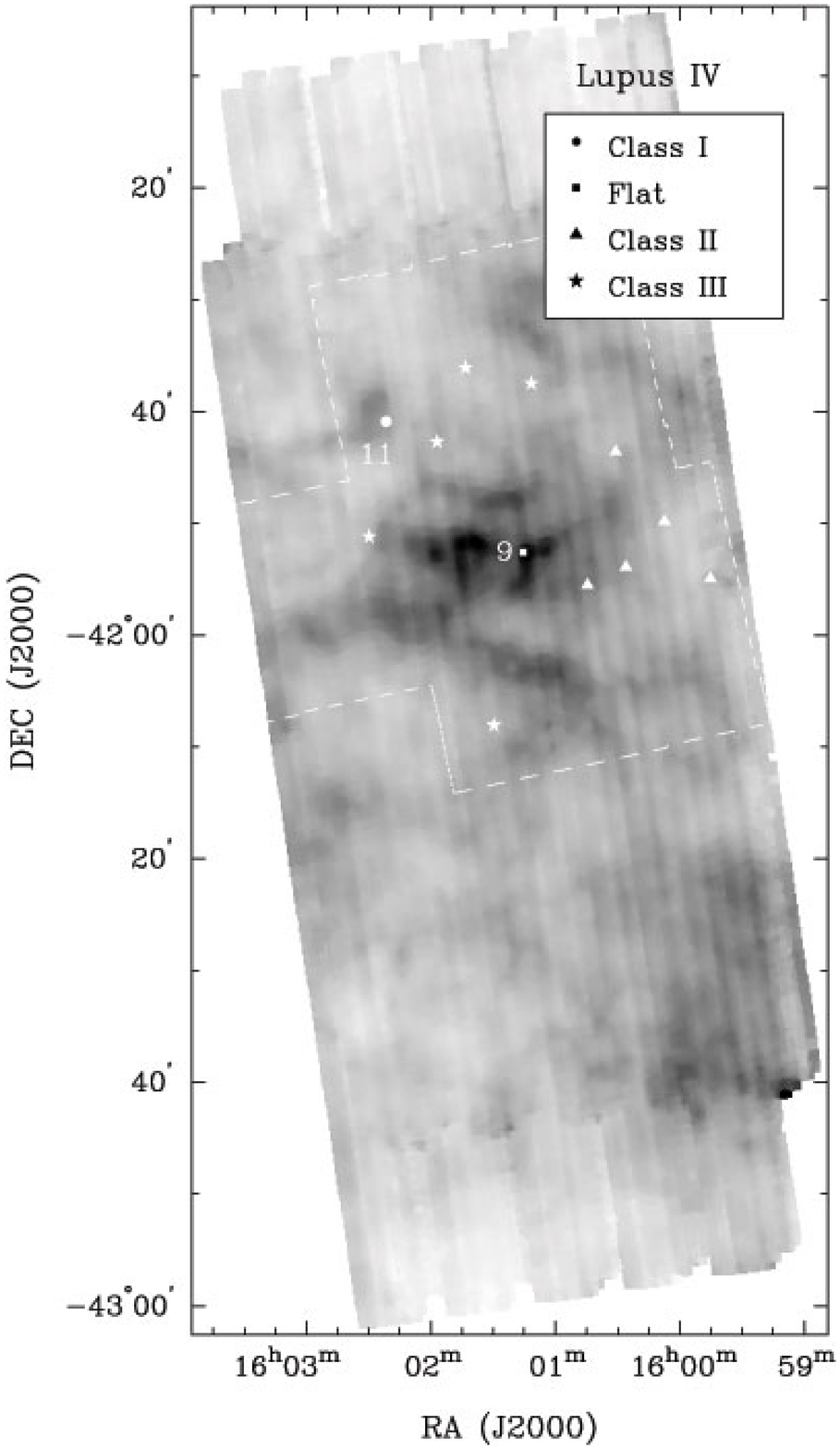}

\caption{\label{fig:lup4-yso}Lupus IV `YSO candidates' overlaid on the 
$160\:\mu$m image. See Figure \ref{fig:lup1-yso} for a description of the 
symbols and their meanings.}

\end{figure}

In Figures \ref{fig:lup1-yso} - \ref{fig:lup4-yso} we have plotted the positions
of all our `YSO candidates' on the $160\:\mu$m emission. Since we require
both IRAC and MIPS data to classify `YSO candidates', we have shown the extent
of the IRAC observations with white outlines.  A few sources have been numbered;
these will be discussed in the following section.

\subsection{Extinction Maps}
\label{sec:avmaps}

\begin{figure}
\plotone{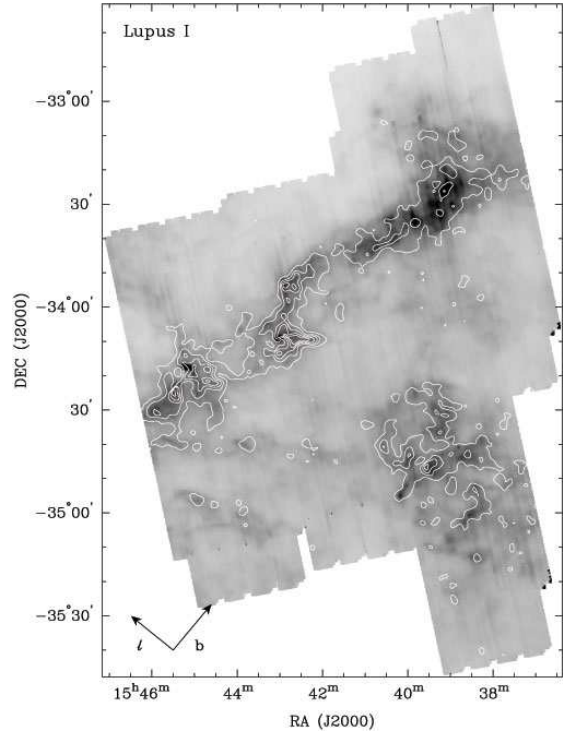}

\caption{\label{fig:lup1-av} The magnitudes of visual extinction (contours) are
overlaid on the $160\:\mu$m emission map for Lupus I. The contours range from 3
to 15 $A_V$, in steps of 3 $A_V$. The extinction map has 2 arcminute resolution.
We have also shown the direction of increasing Galactic $l$ and b coordinates.}

\end{figure}

\begin{figure}
\plotone{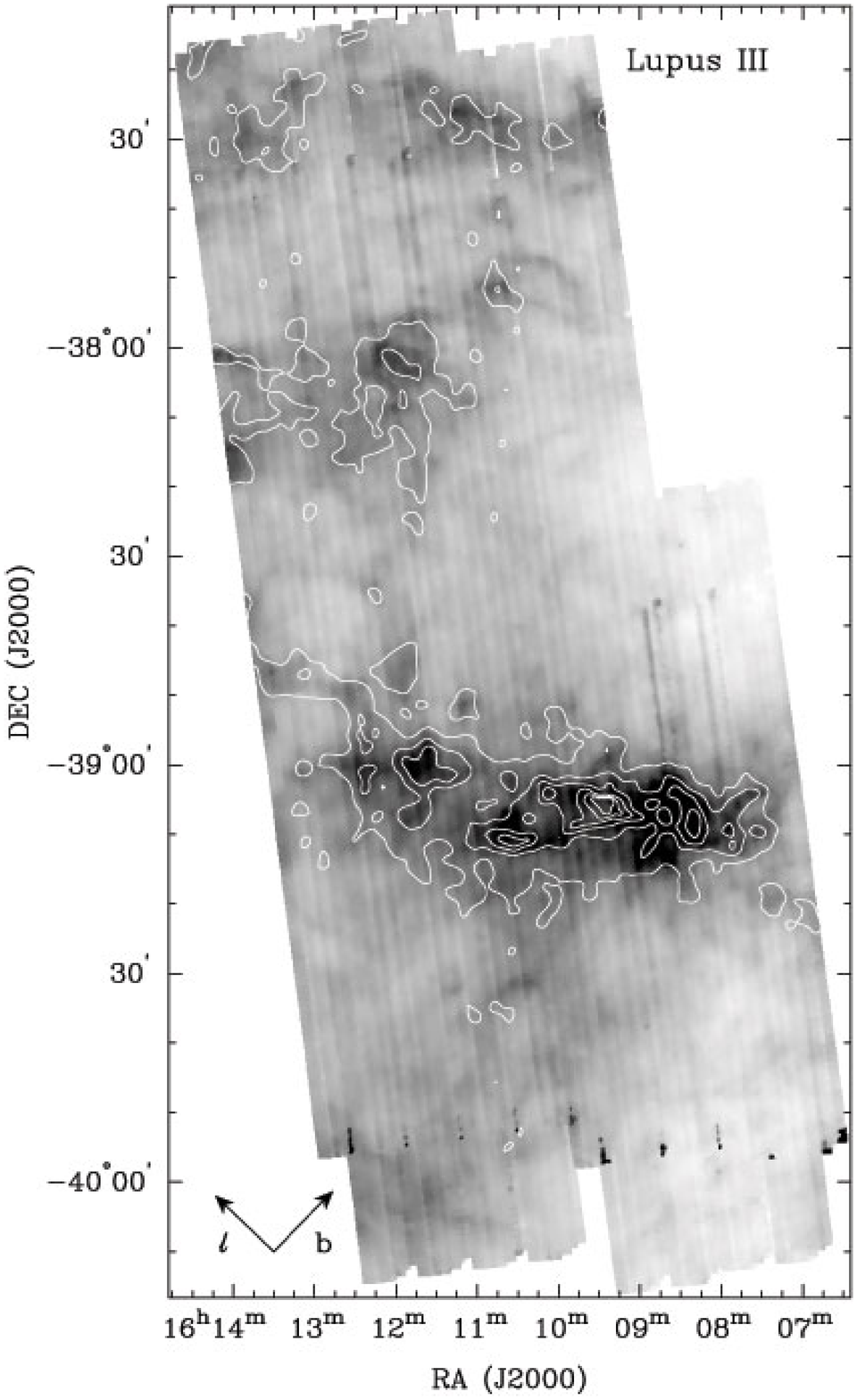}

\caption{\label{fig:lup3-av} The magnitudes of visual extinction (contours) are
overlaid on the $160\:\mu$m emission map for Lupus III. The contours range from
3 to 18 $A_V$, in steps of 3 $A_V$. The extinction map has 2 arcminute
resolution. We have also shown the direction of increasing Galactic $l$ and b
coordinates.}

\end{figure}

\begin{figure}
\plotone{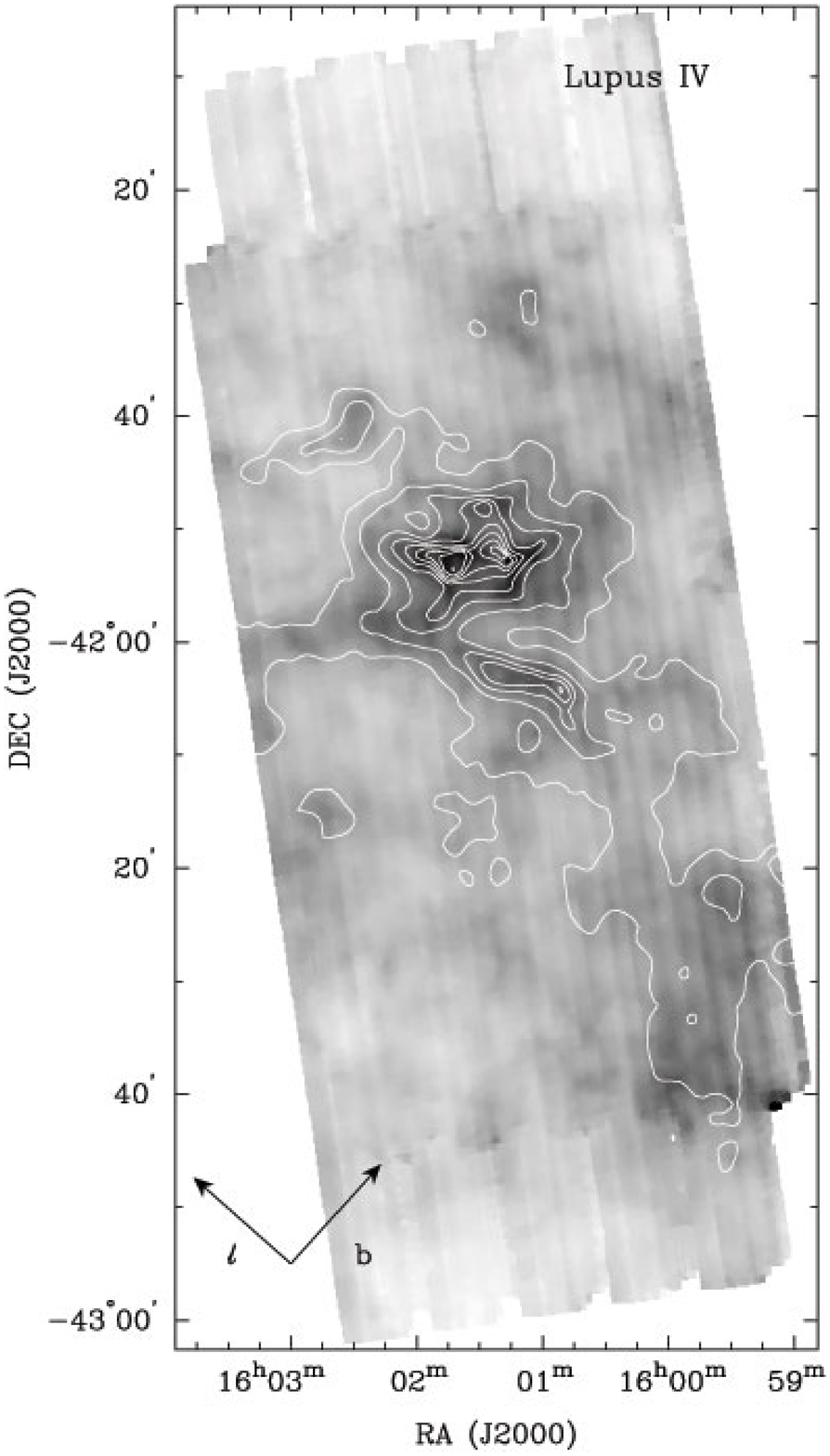}

\caption{\label{fig:lup4-av} The magnitudes of visual extinction (contours) are
overlaid on the $160\:\mu$m emission map for Lupus IV. The contours range from
3 to 27 $A_V$, in steps of 3 $A_V$. The extinction map has 2 arcminute
resolution. We have also shown the direction of increasing Galactic $l$ and b
coordinates.}

\end{figure}

To create extinction maps for our observed regions, we started by excluding
2MASS sources with only an upper limit flux in any of the $JHK_s$ bands. For the
remainder of the 2MASS sources, we calculated $A_V$ values from the $JHK_s$
bands using the method outlined in \S\,\ref{sec:class}. Our method gave us
pinpoint measurements of the $A_V$ along each source's line-of-sight. We
converted these randomly distributed extinction values into uniform maps by
creating a $0\farcm4$ pixel grid on top of the data. At each grid position, the
extinction value in that cell equals the weighted average of the extinctions of
individual points within a $2'$ radius. This average is weighted both by
uncertainty in the individual $A_V$ values and by distance from the center of
the cell with a Gaussian function having a full-width half-maximum (FWHM) equal
to $2'$. Thus, our maps have $2'$ resolution with 5 pixels across the FWHM. We
chose this cell size and resolution to have a well-sampled PSF and to ensure
that our maps would have an extinction value for every pixel, even in the
highest $A_V$ regions. The median number of $A_V$ values in each cell is roughly
40 for Lupus I, and 80 for Lupus III and IV. The percentage of cells with less
than 10 $A_V$ values is $1-2$\% for all three cloud regions. The extinction
contours are shown in Figures \ref{fig:lup1-av}, \ref{fig:lup3-av},
\ref{fig:lup4-av} overlaid on the $160\:\mu$m emission (grayscale). We
subtracted the median off-cloud $A_V$, 0.7 mag., from each map. In all three
maps, the extinction contours start at 3 magnitudes of $A_V$ and increase in
steps of 3 mags. There is a strong correspondence between the $160\:\mu$m
emission and the $A_V$ contours suggesting the $160\:\mu$m emission traces the
dust. A detailed comparison of $A_V$ and $160\:\mu$m emission is beyond the
scope of this paper.

The mass in each Lupus region was estimated by converting the extinction values
in each cell to column density using the equation from \citet{bohlin78},
assuming $R_V = A_V/E(B-V) = 3.1$:

\begin{equation}
\label{eq:nh2av}
N_{HI}/A_V = 1.87 \times10^{21}\:\rm{cm}^{2}\:\rm{mag}^{-1}
\end{equation}

Note that this relationship was established for the diffuse ISM ($R_V = 3.1$)
and not in dense molecular clouds such as we observed, so this equation may not
be valid. This is an important concern since we have used the Weingartner \&
Draine $R_V = 5.5$ dust model in this paper (see \S\,\ref{sec:class}). The data
compiled by \citet{kim96} are consistent with $N_{HI}/A_V$ being independent of
$R_V$, albeit with large uncertainties. In this paper, we will use the value of
$N_{HI}/A_V$ given in Equation \ref{eq:nh2av}, but note that this is possible
source of error in our final mass estimates.

With this equation, plus the molecular weight per hydrogen molecule, $\mu_{H_2}
= 2.8$ \citep[calculated from][]{cox00}, the mass of a hydrogen atom, and the
distance to each Lupus region, we computed masses for each Lupus region from the
extinction maps. In Table \ref{tab:lup-pars} we list the derived mass in regions
with $A_V \ge 3$ for each cloud based on our extinction maps.

It is difficult to compare our mass estimates with values in the literature
since the mapped areas are not always comparable in size.  However, two authors
have observed comparable regions in \mbox{$^{13}$CO ($J = 1 \rightarrow 0$)} and
\mbox{C$^{18}$O ($J = 1 \rightarrow 0$)} (hereafter $^{13}$CO and C$^{18}$O,
respectively.)  \citet{tachihara96} observed Lupus I and III in $^{13}$CO with
$8\arcmin$ resolution and \citet{hara99} observed Lupus I, III, and IV in
C$^{18}$O with a resolution of $2\farcm6$.  Both papers computed cloud masses
from emission $\ge 3\sigma$. With the exception of Lupus I, where these authors'
maps extend about 1 degree further to the east, these $\ge 3\sigma$ areas
are similar to our $A_V \ge 3$ regions. For purposes of comparison to
our results, we will scale these authors' mass estimates using our assumed
distances to the Lupus clouds. This only involves changing their mass estimates
for Lupus III. We have assumed a distance of 200 pc to this cloud whereas
\citet{tachihara96} and \citet{hara99} assumed a distance of $\sim150$ pc for
all Lupus regions.

The $^{13}$CO mass in Lupus I is $1200\:M_\odot$ which is $\sim3\times$ larger
than our extinction mass of $440\:M_\odot$. However, their mapped area is larger
than ours and the $3\sigma$ $^{13}$CO contours cover more area than our $A_V \ge
3$ maps. For Lupus III, the $^{13}$CO mass is $530\:M_\odot$, very similar to
our extinction mass of $690\:M_\odot$. In Lupus I and IV, the C$^{18}$O masses
agree well with our $A_V$ values; the C$^{18}$O masses are 327 and 216
$\:M_\odot$, respectively, for Lupus I and IV compared to 440 and 250
$\:M_\odot$ we derived from $A_V$. The only difference is in Lupus III, where
our $A_V$ mass, $690\:M_\odot$, is $3.5\times$ that derived from C$^{18}$O
($187\:M_\odot$).

Although some differences between our $A_V$ mass estimates and those derived
from $^{13}$CO and C$^{18}$O exist, there are several possible sources of error
that must be considered. Our extinction maps do not cover the exact same areas
as the molecular cloud maps which can potentially lead to large differences in
mass estimates, as is evident in Lupus I. Furthermore, the molecular masses were
estimated assuming a constant excitation temperature of $13$ K, even though
measurements of the excitation temperature in individual dense cores in Lupus
shown variations \citep{vilas00}. Similarly, our $A_V$ masses assumed a constant
grain composition throughout the clouds. This quantity has also been shown to
vary along different sightlines \citep[e.g.][]{cardelli89}. Furthermore, our
$N_{HI}/A_V$ ratio may also be incorrect. A detailed study is needed before
making rigorous comparisons between cloud masses derived using various molecular
tracers and extinction measurements.

\subsection{Selected Sources}
\label{sec:interesting}

\begin{figure}
\plotone{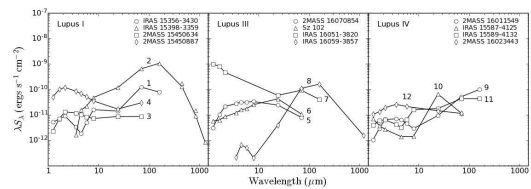}

\caption{\label{fig:interesting} SEDs of selected sources from the three cloud
regions. See \S\,\ref{sec:interesting} for a full discussion of these sources,
and Tables \ref{tab:interesting} and \ref{tab:flux} for the positions and fluxes
of the sources. We have included sub-millimeter and millimeter wavelength data
from the literature for IRAS 15398-3359 and IRAS 16059-3857. For IRAS
15398-3359, the data points at 450 and 850 $\mu$m are from \citet{shirley00}
with $40\arcsec$ (lower points) and $120\arcsec$ (upper) apertures. The 1.3 mm
point comes from \citet{reipurth93} with a $23\arcsec$ aperture.  The 1.2 mm
data point for IRAS 16059-3857 comes from \citet{tachihara07}.}

\end{figure}

In this section we highlight a few sources we selected from Figures
\ref{fig:lupus-cm} and \ref{fig:lupus-cm-m1m2} that are red relative to the
general `YSO candidates' and hence likely to be younger YSOs. From Figure
\ref{fig:lupus-cm} we selected all Flat and Class I sources bright enough in
$K_s$ to not overlap the shaded SWIRE contours denoting `background galaxies'.
Furthermore, we selected sources in Figure \ref{fig:lupus-cm-m1m2} with $[24] -
[70] \geq 4$ and also not overlapping the shaded SWIRE contours. Our final
sample consists of 12 objects. These sources are listed in Tables
\ref{tab:interesting} and \ref{tab:flux} with positions, fluxes, index numbers,
and YSO classification based on $\alpha$. These index numbers are also shown in
the color-magnitude plots (Figures \ref{fig:lupus-cm} -
\ref{fig:lupus-cm-m1m2}), in Figures \ref{fig:lup1-yso} - \ref{fig:lup4-yso},
showing the location of all `YSO candidate' sources, and in Figure
\ref{fig:interesting}, where the SEDs for these objects are plotted.

\subsubsection{Lupus I Sources}

We selected four sources in Lupus I: one previously known and three objects
without references in the literature. The first object is IRAS 15356-3430. The
relative brightness of this source in $K_s$ places it above the typical
extragalactic region in Figure \ref{fig:lupus-cm}. This source is obviously
extended in the 2MASS and IRAC bands. Furthermore, it is located in a relatively
low extinction region to the south instead of near other known protostars. For
these reasons, we believe it to be a background galaxy.

IRAS 15398-3359 is a well-known protostar located in the core B228 \citep[see,
e.g.][]{heyer89,shirley00,shirley02}. It has a molecular outflow which was
discovered in $^{12}$CO ($J = 2 \rightarrow 1$) by \citet{tachihara96}. There is
some correspondence between the Herbig-Haro object HH 185 and the molecular
outflow: the blue-shifted velocity lobe from the molecular outflow overlaps the
blue-shifted [\ion{S}{2}] line from \citet{heyer89}. The [\ion{S}{2}] line is
expected to be excited by the shocks arising from the outflow jet interacting
with the ambient medium. In the IRAC images, the protostar appears extended in
the same direction as the molecular outflow.

The SED of IRAS 15398-3359 is shown in Figure \ref{fig:interesting}. The data
points at 450 and 850 $\mu$m come from \citet{shirley00} using their $40\arcsec$
(lower data points) and $120\arcsec$ (upper points) aperture values, while the
1.3 mm point comes from \citet{reipurth93} with a $23\arcsec$ aperture. This
source does not have a 2MASS detection so it does not appear in Figure
\ref{fig:lupus-cm}.  It is not classified as a `YSO candidate' because 
it is saturated at IRAC1 and IRAC2 and therefore was rejected by our YSO
classification procedure.

To the east of protostar IRAS 15398-3359 is a cluster of objects with infrared
excesses. We have plotted two sources from this region for which we could not
find any references in literature. Both are near Sz 68, a known bright T-Tauri
star. 2MASS 15450887-3417333 is $\sim50\arcsec$ away while 2MASS
15450634-3417378 is about $\sim80\arcsec$ away.

\subsubsection{Lupus III Sources}

Lupus III contains the most active star forming region in the Lupus cloud
complex. HR 5999 dominates the bright nebulous region in the south of our
observed region. This source is a well-known Herbig Ae/Be star. Because this
source is saturated at $3.6$, $4.5$, and $24\:\mu$m wavelengths, we will not
discuss it here. Three of our four sources are nearby, making HR 5999 useful as
a location reference.

2MASS 16070854-3914075 exhibits a fairly flat SED indicating a strong IR excess
over that of a stellar photosphere. This source is located about $15\arcmin$ to
the west of HR 5999. There are several known YSOs in this region, but this
source has not been noted in the literature.

Sz 102 is a well-known T-Tauri star located just to the north of the nebulous
region containing HR 5999. This source is also known as Th 28 or Krautter's
star. \citet{krautter86} discovered an HH outflow from this source (HH 228).
There is some evidence for rotation of the jet \citep{coffey04}. The SED rises
from $1.25-24\:\mu$m, then drops at $70\:\mu$m. This source has also been
observed by the c2d program with the Infrared Spectrometer (IRS) from
$\sim5-35\:\mu$m \citep{jackie06}. The faintness of the object and the
orientation of the jet suggest that the source may have an edge-on disk
\citep{hughes94}. However, the c2d photometry and spectroscopy for this source
do not clearly show the double-peaked structure that would be expected for an
edge-on disk \citep[e.g.][]{wood02}. \citet{stapelfeldt06} observed this source
with HST and did not see any evidence for an edge-on disk.

Our third source is IRAS 16051-3820. This object is outside of the observed IRAC
area, so therefore it cannot be classified as a `YSO candidate'. This source does
have red $K_s - [24]$ and $[24] - [70]$ colors. In Figure
\ref{fig:lupus-cm-m1m2}, this is source \#7 and is located between the Late
Class I and Class II YSO models. In the 2MASS images, two objects, separated
from each other by $\sim5\arcsec$ are clearly visible. These two objects are not
resolved at 24 or $70\:\mu$m.  This source has not been previously identified
as a `YSO candidate' in the literature.

Lastly, IRAS 16059-3857 is located just to the east of the bright nebulous
region. \citet{nakajima03} observed this source with $JHK_s$ imaging and
discovered the region appears as a reflection nebula. They detected an elongated
jet-like structure in $K_s$ and a fan-shaped structure reminiscent of a cavity
created by an outflow. The Herbig-Haro object HH 78 is just to the west. In the
IRAC images, this source appears extended in the roughly the same direction as
the fan-shaped structure seen by \citet{nakajima03}.  \citet{tachihara07} found
this source to be embedded in an H$^{13}$CO$^{+}$ core and measured a 1.2 mm
continuum PSF flux for it.  Because there is no 2MASS data for this source, it
does not appear in Figure \ref{fig:lupus-cm}. However, it is reddest source in
Figure \ref{fig:lupus-cm-m1m2}.

\subsubsection{Lupus IV Sources}

Compared to Lupus III, Lupus IV is a more quiescent region with only a few
T-Tauri stars, most of them previously known. We focus on the four reddest
objects in Figure \ref{fig:lupus-cm} that are not within the `background galaxy'
region denoted by the SWIRE contours.

2MASS 16011549-4152351 is located near the center of globule filament 17 (GF
17), at approximately the center of the mapped region. This source is
consistently rising at all wavelengths longward of $8\:\mu$m and appears to be a
deeply embedded protostar in the filament. There is some nebulosity associated
with the source at 3.6 and $4.5\:\mu$m.

IRAS 15587-4125 is the reddest source in Figure \ref{fig:lupus-cm}, but has only
a modest excess in Figure \ref{fig:lupus-cm-m1m2}. The SED shows a sharp rise
between $8$ and $24$. This source is the known planetary nebula PN G336.9+08.3
\citep{acker92}.

IRAS 15589-4132 has an SED that rises rapidly at wavelengths longward of
$4.5\:\mu$m. Because this source is extended in all four IRAC bands and in
the 2MASS $JHK_s$ images, we suspect this source to be a galaxy.  This source
has no identification in the literature.

Our last source, 2MASS 16023443-4211294, is not identified as a `YSO candidate'
because it falls outside of the observed area of IRAC2 ($4.5\:\mu$m) and IRAC4
($8\:\mu$m). The only reference in the literature for this source is an x-ray
source about $12\arcsec$ away. The source does not appear extended in 2MASS,
IRAC, or MIPS images. This source is located far from the globular filament, so
it tempting to identify this object as a background galaxy. However, the SED for
this object is fairly flat, which is consistent with other `YSO candidates'
instead of the rising SED found in our suspected galaxies.

\section{Summary}

We presented maps of three regions within the Lupus molecular cloud complex at
24, 70, and $160\:\mu$m observed with MIPS. We discussed the c2d data reduction
pipeline \citep{evans06} and the differences between the c2d pipeline and our
catalogs used here. Our catalogs were bandmerged with 2MASS and SED fitting was
performed to classify objects. After removing asteroids and selecting only the
highest reliability MIPS sources, our final catalogs contain 1790 sources for
Lupus I, 1950 for Lupus III, and 770 for Lupus IV. In addition, we have 
created three c2d processed catalogs from a subset of the ELAIS N1 SWIRE data.
Each catalog was trimmed to have similar sensitivity and extinction as each
of our cloud regions.

We compared the $24\:\mu$m source counts in each region to our c2d processed
SWIRE catalog, a catalog for a nearby off-cloud region, and the star counts
derived from the Wainscoat model. We find that Lupus I, at the highest Galactic
latitude, shows an excess of sources at fainter fluxes over the Wainscoat model,
while III and IV, at more modest latitudes, are consistent with Wainscoat over a
broad range of fluxes. The faint Lupus I counts are consistent with the c2d
processed SWIRE catalog suggesting this region is more dominated by background
galaxies.

Using an empirical method developed by \citet{harvey07b}, we created a sample of
objects that is strongly enriched in YSOs and minimizes contamination by
background galaxies. We identify 103 `YSO candidates' within the three Lupus
regions surveyed. Of these, we find positional matches with 26 of the 40
confirmed classical T-Tauri stars in Lupus and 6 of the suspected protostars
from \citet{lopezmarti05} and \citet{comeron03}. When we break down our `YSO
candidate' sample by $\alpha$, the best-fit slope to the flux from $2-24\:\mu$m,
we find that almost all of the sources fall into the Class II and Class III
categories. This suggests that the most of the protostars within Lupus are older
and more evolved. There are a few of the younger Class I and Flat spectrum
sources in each region.

Our SED fitting procedure also computes the $A_V$ for each object. We have used
the 2MASS data to construct extinction maps of the three regions and compared
them to the $160\:\mu$m emission. Our cloud masses derived from the extinction
maps are largely comparable to previous studies, except for Lupus I, where
previous authors have mapped a larger region than our observations, and in Lupus
III, where our extinction mass is $3.5\times$ larger than the C$^{18}$O mass
from \citet{hara99}. However, there are several different factors that could
alter these mass estimates such as assumptions about temperature or dust
composition. A more detailed study is needed.

Based on two color-magnitude diagrams we selected 12 objects with very red $K_s
- [24]$ and $[24] - [70]$ colors for individual mention. Three of these are
previously known protostars: IRAS 15398-3359, Sz 102, and IRAS 16059-3857. We
suspect two other sources of being bright background galaxies, one object is a
known planetary nebula, and one object is a double star. We are left with five
previously unknown potential protostars. Further observations are needed to
confirm whether these sources are bona-fide YSOs.

\acknowledgments

We would like to thank the staff at the Lorentz Center in Leiden for their
hospitality in hosting several meetings that contributed to this paper. Support
for this work, part of the Spitzer Legacy Science Program, was provided by NASA
through contracts 1224608, 1230782, and 1230779 issued by the Jet Propulsion
Laboratory, California Institute of Technology, under NASA contract 1407.
Astrochemistry in Leiden is supported by a NWO Spinoza grant and a NOVA grant.
KEY was supported by NASA under Grant No. NGT5-50401 issued through the Office
of Space Science. The research of JKJ was supported by NASA Origins Grant
NAG5-13050.  LGM is supported by NASA Origins Grant NAG510611.

This research has made use of the SIMBAD database, operated at CDS, Strasbourg,
France. This research has made use of the NASA/IPAC Extragalactic Database (NED)
which is operated by the Jet Propulsion Laboratory, California Institute of
Technology, under contract with the National Aeronautics and Space
Administration.

\end{document}